\documentclass[aps,showpacs,superscriptaddress]{revtex4}
\usepackage[dvips]{graphicx}
\usepackage{epsfig,rotating}
\usepackage{amsmath,amssymb}
\numberwithin{equation}{section}

\newcommand{\be}{\begin{equation}}
\newcommand{\ee}{\end{equation}}
\newcommand{\bea}{\begin{eqnarray}}
\newcommand{\eea}{\end{eqnarray}}
\begin{document}
\title{Dynamical renormalization group approach to relaxation in
  quantum field theory}
\author{D. Boyanovsky}
\email{boyan@pitt.edu} \affiliation{Department of Physics and
Astronomy, University of Pittsburgh, Pittsburgh, Pennsylvania
15260, USA}\affiliation{LPTHE, Universit\'e Pierre et Marie Curie
(Paris VI) et Denis Diderot (Paris VII), Tour 16, 1er. \'etage, 4,
Place Jussieu, 75252 Paris, Cedex 05, France}
\author{H. J. de Vega}
\email{devega@lpthe.jussieu.fr} \affiliation{LPTHE, Universit\'e
Pierre et Marie Curie (Paris VI) et Denis Diderot (Paris VII),
Tour 16, 1er. \'etage, 4, Place Jussieu, 75252 Paris, Cedex 05,
France}\affiliation{Department of Physics and Astronomy,
University of Pittsburgh, Pittsburgh, Pennsylvania 15260, USA}

\date{\today}

\begin{abstract}
 The real time evolution and relaxation of expectation values of quantum fields
and of quantum states are computed as initial value problems by
implementing the dynamical renormalization group (DRG). Linear
response is invoked to set up the renormalized initial value
problem to study  the dynamics of the expectation value of quantum
fields. The perturbative solution of the equations of motion for
the field expectation values of quantum fields as well as the
evolution of quantum states features {\bf secular terms}, namely
terms that grow in time and invalidate the perturbative expansion
for late times. The DRG provides a consistent framework to {\bf
resum} these secular terms and yields a uniform asymptotic
expansion at long times. Several relevant cases are studied in
detail, including those of threshold infrared divergences  which
appear in gauge theories at finite temperature and lead to
anomalous relaxation. In these cases the DRG is shown to provide a
resummation akin to Bloch-Nordsieck but directly in real time and that
goes beyond the scope of Bloch-Nordsieck and Dyson resummations.
The nature of the resummation program is discussed in several
examples. The DRG provides a framework that is consistent,
systematic and easy to implement to study the non-equilibrium
relaxational dynamics directly in real time that does not rely on
the concept of quasiparticle widths.
\end{abstract}


\maketitle

\tableofcontents

\section{\label{sec:intro} Introduction}
The study of relaxation phenomena in quantum field theory or many
body systems out of equilibrium is a timely topic of
multidisciplinary interest with applications ranging from
ultrarelativistic heavy ion collisions, to cosmology and condensed
matter physics. The program of ultrarelativistic heavy ion
collisions focuses on creating a new state of matter, the
Quark-Gluon plasma\cite{QGP}-\cite{elze} which existed about one
microsecond after the Big Bang. At the ultrarelativistic heavy ion
colliders RHIC at Brookhaven and the forthcoming LHC at CERN, this
new state is conjectured to be formed and to exist for about
$10^{-22}$ seconds. Thus probing the physics of this short lived
transient state experimentally requires the theoretical
understanding of  non-equilibrium phenomena on unprecedented short
time scales.

In cosmology the challenge of understanding non-equilibrium
phenomena leading to thermalization and equilibration, phase
transitions and baryogenesis\cite{kolb,lyth,riotto,chalonge}
requires a novel set of tools to study relaxation in extreme
environments and on very short time scales.

In condensed matter, transport phenomena in mesoscopic
systems\cite{schon} an ultrafast spectroscopy in
semiconductors\cite{ultrafast,qk} explore relaxation of
non-equilibrium states on short time and distance scales.

The usual tools to study relaxation and transport phenomena is
based on kinetic equations which describe the dynamics of single
(quasi) particle distribution functions or the concept of the
damping rate of quasiparticles which is extracted from the
imaginary part of the single particle self-energy and is
associated with exponential decay.

In gauge theories at high temperature it has been found that
charged particles of hard momentum feature infrared singularities
on their particle mass shells that lead to anomalous damping
rates\cite{pisa}. These infrared singularities are  a consequence
of the emission and absorption of transverse photons which are
only dynamically screened by Landau damping.

A study of the relaxation of charged excitations directly in real
time by means of a Bloch-Nordsieck resummation of the
propagator\cite{blaizot} revealed that this anomalous damping rate
actually implies non-exponential relaxation.

A novel approach, the dynamical renormalization group (DRG) was
recently introduced\cite{drg1,drgqed}. It leads to a study of
relaxation and kinetics directly in real time and reveals the
non-exponential relaxation of charged particles in high
temperature gauge theories.

Th DRG approach to relaxation and kinetics borrows the main
concepts from the study of the asymptotics in ordinary
differential equations\cite{goldenfeld,kuni,otrosdrg}.

The dynamical renormalization group (DRG) has been recently
implemented to study transport phenomena in high temperature gauge
theories: in particular the electrical conductivity has been
extracted in high temperature quantum
electrodynamics\cite{conductivity}. This approach to studying
relaxation and transport phenomena directly in real time does not
rely on the quasiparticle concept and instead extracts the real
time dynamics from a resummation program that is akin to the
renormalization group in critical phenomena and deep inelastic
scattering.

In this article we present the main concepts of the DRG explicitly
obtaining the relaxation of expectation values in quantum field
theory as well as the time evolution of quantum states.

In section II and III we begin by setting up the fully
renormalized initial value problem that describes the real time
evolution of the expectation value of bosonic quantum field out
equilibrium in linear response. In section IV we solve the real
time evolution equations in several relevant cases by Laplace
transform and highlight the case in which the emission and
absorption of massless quanta introduce threshold infrared
divergences. These divergences render the concept of a
quasiparticle decay rate ambiguous and show that a Bloch-Nordsieck
resummation of these infrared divergences lead to power law
relaxation at zero temperature. At finite temperature the infrared
divergences are more severe and a Bloch-Nordsieck resummation of
the propagator is not readily available.

In section V we  introduce the dynamical renormalization group
resummation and compare to the results obtained by the Laplace
transform solution highlighting the relevant resummations. This
method allows to treat the most infrared severe cases of threshold
divergences in a similar simple manner thus providing a very
powerful resummation scheme.

In section VI we present a scalar theory model that features the
most relevant cases: stable particles and resonances and different
cases of threshold infrared divergences which are common to gauge
theories.

In section VII  the time evolution of states in quantum mechanics
is solved in the DRG approach and a direct comparison with
standard results on energy shifts and decay rates is established.

In section VIII  the resummations implied by the dynamical
renormalization group,  and the issue of separation of time scales
are discussed in detail. Section IX summarizes the conclusions.

Several appendices contain relevant mathematical  details and a
simple albeit illuminating example.

\section{From linear response to an initial value problem:}
We are  interested in studying the real time evolution of
expectation values of  field operators. Consider a scalar field
theory with an interacting Lagrangian density ${\cal L}[\Phi] $
the expectation value of the scalar field $\Phi$ can be obtained
from linear response to an external c-number source term $J$. The
appropriate formulation of real time, non-equilibrium dynamics is
that of Schwinger-Keldysh\cite{ctp}-\cite{tadpole1} in which a
path integral along a contour in imaginary time is required to
generate all of the non-equilibrium Green's functions. Along the
contour the fields carry labels $\pm$ corresponding to the forward
branch ($+$) along which the time variable takes values $-\infty
\leq t \leq \infty$, and the backward branch ($-$) along which $t$
runs back from $\infty$ to $-\infty$. The two branches describe
the time evolution of a density matrix which requires the time
evolution forward and backward in time (for details see
refs.\cite{ctp,disip,tadpole1}).

The non-equilibrium Lagrangian density along this contour is
therefore given by\cite{ctp}-\cite{tadpole1},
\begin{equation}
{\cal L}_{NEQ}[\Phi^+,\Phi^-;J] = {\cal L}[\Phi^+]+J \; \Phi^+ - {\cal
L}[\Phi^-]-J \; \Phi^- \; . \label{lagraneq}
\end{equation}
The non-equilibrium expectation value of the scalar field in a
linear response analysis is given by
\begin{equation}
\langle \Phi^+(\vec x,t) \rangle   =  \langle \Phi^-(\vec x,t)
\rangle = \phi(\vec x,t) = \int_{-\infty}^{+\infty} d^3{\vec x}'
\; dt' \;  G_R(\vec x- {\vec x}',t-t') \;  J({\vec x}',t') \;
,\label{linresp}
\end{equation}
with the retarded Green's function
\begin{eqnarray}
&&G_R(\vec x- {\vec x}',t-t')   =   \left[G^>(\vec x- {\vec
    x}',t-t')-G^<(\vec x- {\vec x}',t-t')\right]\Theta(t-t') =  i
\langle \left[\Phi(\vec x,t), \Phi( {\vec x}',t' \right]\rangle
\Theta(t-t')  \; , \label{retcomm}
\end{eqnarray}
where the expectation value is in the full interacting theory but
with vanishing source. Consider an external source term that is
adiabatically switched on in time from $t \rightarrow -\infty$ and
of the form,
\begin{equation}
J({\vec x}',t') = J({\vec x}') \;  e^{\epsilon t'} \;  \Theta(-t') \quad
, \quad \epsilon \rightarrow 0^+  \; .\label{source}
\end{equation}
The retarded nature of $G_R(\vec x- {\vec x}',t-t')$ results in
that, \be\label{inival} \phi(\vec x,t=0)  =  \phi_0(\vec x) \quad
, \quad \dot{\phi}(\vec x,t<0) =  0 \; , \ee where $ \phi_0(\vec
x) $ is determined by $J(\vec x)$ (or vice versa), and the second
line follows in the limit $\epsilon \rightarrow 0$. For a given
initial value $ \phi_0(\vec x) $ the current  $J(\vec x)$) can be
found from from eq.(\ref{linresp}) and the vanishing of the
derivative for $t<0$ is a consequence of the retarded nature of
$G_R$.

The linear response problem with the initial conditions at $t=0$
given by eq.(\ref{inival}) can now be turned into an initial value
problem for the {\em equation of motion}
 of the expectation value by using the (integro-) differential operator
${\cal O}_{(\vec x,t)}$ inverse of $G_R(\vec x- {\vec x}',t-t')$
\begin{equation}
{\cal O}_{(\vec x,t)}\phi(\vec x,t)  =   J(\vec x,t) \quad , \quad
\phi(\vec x,t=0)  =  \phi_0(\vec x) \quad , \quad \dot{\phi}(\vec
x,t<0) =  0  \; , \label{equlin}
\end{equation}
for the source term given by eq.(\ref{source}). Within the
non-equilibrium formulation, the equation of motion of the
expectation value is obtained via the tadpole
method\cite{disip,tadpole1} and automatically leads to a retarded
initial value problem by coupling an external source that
satisfies eq.(\ref{source}).

\section{A  scalar theory}

Although the results obtained below are generic for bosonic,
fermionic or gauge theories, we will focus our analysis on the
interacting bosonic theory to highlight  the main ideas in a
definite setting. All of the steps can be generalized to more
general theories, either fermionic, bosonic and or gauge theories.

We now establish the general formulation for an arbitrary bosonic
scalar field theory, deferring to specific examples in a later
section. The type of theories under consideration are generally of
the form
\begin{equation}
{\cal L} = \frac{1}{2}(\partial_{\mu}\Phi)^2 -\frac{1}{2}\;
m^2_{0} \; \Phi^2 +J_0 \;
\Phi+\mathcal{L}_{int}\label{lagrascalarctp} \; ,
\end{equation}
\noindent where the source coupled to the  field $\Phi$ has been
introduced to provide an initial value problem for its expectation
value  as explained in the previous section.

The fully renormalized equations of motion are obtained by
introducing the usual wave function, mass and coupling
renormalizations,
\begin{eqnarray}
&&\Phi= Z^{\frac{1}{2}} \; \Phi_{r} \quad ,  \quad
J_r=Z^{\frac{1}{2}} \; J_0 \quad ,  \quad m^2_{0} \;  Z = m^2_{r}
\;  Z+ \delta m^2  \; , \label{renorma}
\end{eqnarray}
\noindent with a renormalization of the interaction terms included
in the interaction Lagrangian.

 We now suppress the label $r$ with all quantities being renormalized,
 and write the Lagrangian density in terms of renormalized quantities
and counterterms
\begin{eqnarray}\label{contra}
 {\cal L} & = & \frac{1}{2}(\partial_{\mu}\Phi)^2 -\frac{1}{2}\; m^2 \; \Phi^2
  +J \; \Phi+ {\cal L}_{ct}+ \mathcal{L}_{int} \nonumber \\
 {\cal L}_{ct} &  = &  \frac{1}{2}\left[(Z-1) \; (\partial_{\mu}\Phi)^2-
m^2 \; \Phi^2\right] -\frac{1}{2} \; \delta m^2 \; \Phi^2 \; .
\end{eqnarray}
 The counterterms are adjusted in perturbation theory as usual, and
  the interaction Lagrangian is in terms of renormalized quantities
  and the counterterms associated with the interaction.

Writing,
$$\Phi(\vec x,t) = \psi(\vec x,t)+\phi(\vec x,t) \quad \mbox{with}\quad
\langle \psi(\vec x,t) \rangle =0\quad ; \quad  \phi(\vec x,t)=
\langle \Phi(\vec x,t) \rangle \; , $$
 using the tadpole condition\cite{disip,tadpole1}
\begin{equation}\label{tadpole}
\langle \psi(\vec x,t) \rangle = 0 \; ,
\end{equation}
\noindent  and taking spatial Fourier transforms,  we find the
following equation of motion linear in the amplitude for the
expectation value for $t>0$ (when the external source $J$
vanishes)
\begin{equation}
\ddot{\phi}_k(t)+ \omega^2_k \; \phi_k(t) +
\int_{-\infty}^{+\infty} \Sigma^R_k(t-t') \; \phi_k(t') \; dt' = 0
\; , \label{eqnofmotionscalar}
\end{equation}
with $\omega^2_k=k^2+m^2$, and used that the adiabatic current
vanishes for $t>0$ [see eq.(\ref{source})]. Here $ \Sigma^R_k(t-t') $
stands for the renormalized retarded self-energy. That is, $
\Sigma^R_k(t-t') $ is obtained from the bare one after mass and
wavefunction renormalization.

As discussed in the previous section, the
 source is chosen so that $\phi_k(t=0)=\phi_k(0)\; ; \;
 \dot{\phi}_k(t\leq 0) =0$. $\Sigma^R_k(t-t')$ is the retarded
 self-energy which is obtained systematically in perturbation
 theory. The Fourier representation of  the retarded self-energy is given by,
\begin{equation}
\Sigma^R_k(t) = \int^{+\infty}_{-\infty} \frac{d\omega}{2\pi} \;
e^{-i
  \,\omega \,   t} \; {\Sigma}_k(\omega) \; ,
\end{equation}
It is very useful to express the self-energy in a spectral
representation. In renormalizable field theories, usually one needs
to make two subtractions in the dispersion relation. However, in the
scalar models considered below, the wave function renormalization is finite and
one subtraction is enough (we can set $Z=1$). We can then write a spectral
representation in the form,
\begin{equation}\label{disprel}
\Sigma_k(\omega+i0) - \Sigma_k(0) = \omega \int^{+\infty}_{-\infty} d\omega' \;
\frac{\rho_k(\omega')}{\omega' \; (\omega-\omega' + i0)} \; .
\end{equation}
\noindent where the spectral density $\rho_k(\omega)$ for real
bosonic fields is an odd function of $\omega$,
\be\label{specrepre}
\rho_k(-\omega) =  -\rho_k(\omega)  \; .
\ee
Eq.(\ref{disprel})
implies that $\rho_k(\omega)$ is positive for positive $\omega$. 

The real and imaginary parts of ${\Sigma}_k(\omega )$ are given by
\begin{eqnarray}\label{realpart}
\mathrm{Re}\left[ \Sigma_k(\omega) - \Sigma_k(0)\right] &  = &\omega \; 
 \mathcal{P}\int^{+\infty}_{-\infty} d\omega'
 \; \frac{\rho_k(\omega)}{\omega' \; (\omega -\omega')} = 2 \,\omega^2 \;
 \mathcal{P}\int^{+\infty}_{0} d\omega'
 \; \frac{\rho_k(\omega')}{\omega' \; (\omega^2 -\omega'^2)} \quad , \quad
\mathrm{Im}{\Sigma}_k(\omega)  =   -\pi \; \rho_k(\omega) \; ,
\end{eqnarray}
\noindent where $\mathcal{P}$ stands for the principal part.

The retarded self-energy can thus be written in a spectral
representation in the form 
 \be
\Sigma^R_k(t)  =  -i \, \Theta(t) \int^{+\infty}_{-\infty}
\rho_k(\omega) \; e^{-i\omega \, t} \; d\omega = -2  \,
\Theta(t)\int^{+\infty}_0 d\omega \;  \rho_k(\omega) \; \sin\omega \, t 
\; .
 \ee
and the inverse Fourier transformation,
\be\label{trafin}
{\Sigma}_k(\omega) = \int_0^{+\infty} e^{i\omega \, t} \; 
\Sigma^R_k(t) \; dt \quad , \quad \mathrm{Im} \, \omega > 0 \; .
\ee
In what follows we use the usual perturbative expansions in terms
of a dimensionless coupling $ \lambda $ for the self-energy and
the renormalization counterterms
\begin{eqnarray}\label{expansions}
\Sigma_k(\omega) & = & \lambda \;\Sigma^{(1)}_k(\omega)+\lambda^2
\; \Sigma^{(2)}_k(\omega)+{\cal O}(\lambda^3) \quad  , \quad
\rho_k(\omega)  =  \lambda \; \rho^{(1)}_k(\omega)+\lambda^2 \;
\rho^{(2)}_k(\omega)+{\cal O}(\lambda^3) \label{pertrho}  \; .
\end{eqnarray}
The initial value problem with the condition that $\dot{\Phi}(\vec
x,t<0)=0$ can be solved via Laplace transform.

\subsection{The Laplace transform}

Introducing  the Laplace transform of $\phi_k(t)$ as
\begin{equation}\label{laplafield}
\varphi_k(s) = \int_0^{\infty} dt \; e^{-st} \;  \phi_k(t)  \quad ,
\quad \mathrm{Re} \, s > 0 \; . 
\end{equation}
\noindent 
\noindent and that of the retarded self-energy,
\be\label{lapsig}
\widetilde{\Sigma}_k(s)= \int_0^{\infty} dt \; e^{-st} \;\Sigma^R_k(t)
\; ,
\ee
which can be read off eq.(\ref{trafin}) after the analytic
continuation $ i \, \omega = -s $,
\begin{equation}
\widetilde{\Sigma}_k(s=-i\omega+0)={\Sigma}_k(\omega+i0)
= \mathrm{Re}\,{\Sigma}_k(\omega)+i  \;
\mathrm{Im}\,{\Sigma}_k(\omega) \label{equivalence}
\end{equation}
\noindent where the real and imaginary parts of the self-energy
are given by eqs.(\ref{realpart}).  Notice that $s$ approaching
the imaginary axis from the right half-plane, $ s = -i\omega + 0 $,
corresponds to the causal choice $ \omega + i 0 $.

The spectral representation of the Laplace transform $
\widetilde{\Sigma}_k(s) $ follows from eq.(\ref{disprel}) by analytic
continuation, 
\begin{equation}\label{laplasigma}
\widetilde{\Sigma}_k(s) - \widetilde{\Sigma}_k(0)
=  2 \, s^2 \, \int^{+\infty}_{0}
\frac{d\omega}{\omega} \;\frac{\rho_k(\omega)}{\omega^2+s^2} \; ,
\end{equation}
\noindent where we have used the property (\ref{specrepre}) in the
last line and $\widetilde{\Sigma}_k(0) =  \Sigma_k(0) $.

\medskip

Upon taking the Laplace transform,  eq.(\ref{eqnofmotionscalar})
becomes
 the following algebraic equation (see Appendix \ref{sec:appendix1}),
\begin{equation} \label{ecmovks}
[s^2+\omega^2_k+\widetilde{\Sigma}_{ k} (s) ] \; \varphi_k( s)  -s
\, {\phi}_k(0) + \frac{ { \phi}_k(0)}{s} \left[
{\widetilde{\Sigma}}_k (0)- \widetilde{\Sigma}_{ k} (s)\right] = 0
\; .
\end{equation}
The spectral density
is consistently computed in a perturbative expansion in terms of
the renormalized coupling constant. The renormalization aspects
are well known and therefore we do not well into them. We just
emphasize that the initial value problem is fully renormalized
formulated in terms of the fully renormalized self-energy $\Sigma_k(\omega+i0)$
for which the relation (\ref{equivalence})
holds in terms of the fully renormalized $\widetilde{\Sigma}_k(s)$.

 The solution of Eq.(\ref{ecmovks}) is  therefore given by
\begin{equation} \label{sols}
\varphi_k(s) = \frac{ \phi_k( 0)}{s} \left[ 1 -\frac{\omega^2_k+
{\widetilde{\Sigma}}_k(0)}{s^2+\omega^2_k+{\widetilde{\Sigma}}_k
(s)}
  \right] \; .
\end{equation}
This expression manifestly implies that the Laplace transform
method to obtain the real time evolution is equivalent to a Dyson
(geometric) resummation of self-energy insertions.

The real time evolution is obtained from the inverse Laplace
transform given by
\begin{equation} \label{invlap}
{ \phi}_k( t) = \int_{\Gamma} \frac{ds}{2\pi i} \; e^{st}
\;\varphi_k( s) \; .
\end{equation}
where the contour $\Gamma$ runs from $-i \infty$ to $ +i \infty$
parallel to the imaginary axis in the $\mathrm{Re}\,s>0$
half-plane. We note that $s=0$ is \textit{not} a pole since the
residue vanishes identically in eq.(\ref{sols}).

Notice that eq.(\ref{sols}) provides a \textit{ non-perturbative}
solution for the time evolution of the expectation value of the
field for a given \textit{perturbative } expression for $
\widetilde{\Sigma}_{k} (s) $. Eq.(\ref{sols}) is an explicit
non-perturbative resummation of the Schwinger-Dyson series. For a
given order in perturbation theory for the self-energy this result
corresponds to a geometric (Dyson) series with the self-energy
calculated at a given order in perturbation theory.

The information about the initial state is encoded in $
\widetilde{\Sigma}_{ k} (s)  $. The self-energy is computed
perturbatively with propagators containing the information about
the initial state. For example, in the thermal case they depend on
the initial temperature, and  for non-zero density they depend on
the chemical potential.

In order to compute the integral eq.(\ref{invlap}) it is
convenient to deform the contour to the left $s$-plane. The
singularities of $ \widetilde{\Sigma}_{ k} (s) $ are usually along
the imaginary $s$ axis, such as  poles at the particle dispersion
relations and production and thermal cuts.

\section{The time evolution via Laplace transform}

We now study in detail the solution of the time evolution via
Laplace transform in several relevant cases.

\subsection{Isolated particle pole: stable particles}

We begin by studying the simple case in which the spectral density
for $\omega >0$ has support above the position of the particle
pole  $\Omega_k (\sim \omega_k)$, namely
\begin{equation}
\rho_k(\omega) \neq 0 ~\mathrm{for}~ \Omega_k< \omega_{th}<\omega
\leq \infty \; .\label{partpole}
\end{equation}
In this case $ s^2+\omega^2_k+\widetilde{\Sigma}_k(s) $ vanishes
at two  points along the imaginary axis at $s=\pm i\Omega_k$, with
$\Omega_k$ being the real solution of the equation
\begin{equation} \label{polo}
\Omega^2_k -\omega^2_k -\Sigma_{k} (\Omega_k) = 0 \; .
\end{equation}
Since $\rho_k(\Omega_k) =0$ the imaginary part of the self-energy
vanishes at this point. This situation corresponds to a
\emph{stable} particle.

\begin{figure}[ht!]
\includegraphics[width=2.5in,keepaspectratio=true]{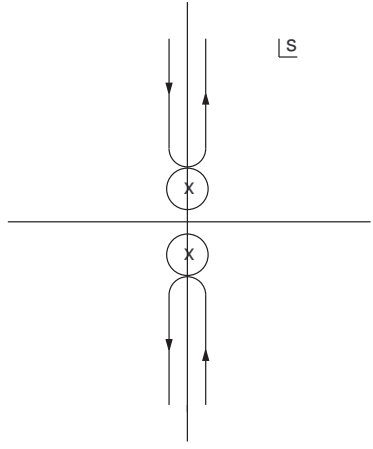}
\caption{Contour for the time evolution.}\label{fig:stable}
\end{figure}

To lowest order in the coupling $\lambda$ we find,
\begin{eqnarray}
\Omega_k & = & \omega_k + \delta \omega_k \label{lopole} \quad ,
\quad
 \delta \omega_k  = \frac{\lambda}{2\omega_k} \;
\mathrm{Re}{\Sigma}^{(1)}_k(\omega_k) \; .
\end{eqnarray}
 The Laplace transform (\ref{invlap}) is
performed by wrapping the contour to the one shown in fig.
\ref{fig:stable} where the cuts along the imaginary axis run along
$ \omega_{th}\leq |\mbox{Im} s | < +\infty $ with $\omega_{th}$
being the particle production threshold.

Deforming the contour $ \Gamma $ to the left yields the integral
as the sum of the pole contributions plus the integral over the
discontinuity of the cut along the imaginary $s$-axis. This cut
may be a production cut for $ | \mbox{Im} \, s| > \omega_{th} $ or
at finite temperature a thermal cut along   $ | Im \, s| < k $
(see section \ref{sec:specexa}) below. In all these cases,
provided the poles are separated from the cuts, which is the case
for stable particles we can write the integral in
eq.(\ref{invlap}) as
\begin{equation}\label{lapexa}
\delta({\vec k}, t) \equiv \frac{{ \phi}_k( t)}{{ \phi}_k( 0) } =
\mathcal{Z}_k\; \cos(\Omega_k t) + \left[
  \omega^2_k+\Sigma_k(0)\right]  \; \int_{-\infty}^{+\infty}
\frac{d\omega}{\omega} \; \frac{\rho_k(\omega) \;
\cos(\omega t)}{\left[  \omega^2-\omega^2_k-
\mathrm{Re}{\Sigma}_k(\omega) \right]^2 +
\left[\mathrm{Im}{\Sigma}_k(\omega)\right]^2} \; .
\end{equation}
where
\begin{equation}\label{poleres}
\mathcal{Z}_k =
\frac{\omega^2_k+\Sigma_k(0)}{\Omega^2_k\left[1+
\frac{1}{2\Omega_k}\frac{\partial \mathrm{Re}
{\Sigma}_k(\omega)}{\partial \omega}\Big|_{\omega=\Omega_k}
\right]}
\end{equation}
To lowest order in the coupling $\lambda$ we find,
\begin{equation}\label{Zlo}
\mathcal{Z}_k = 1 - \lambda  \int_{-\infty}^{+\infty} d\omega \;
\frac{\rho^{(1)}_k(\omega)}{\omega} \;  \mathcal{P}
\frac{\omega^2_k}{(\omega^2-\omega^2_k)^2} \; ,
\end{equation}
\noindent this wave function renormalization is \emph{finite}
irrespective of the choice of renormalization counterterm $Z$ even
in renormalizable theories where $\rho_k(\omega) \sim \omega^2$ as
$\omega \rightarrow \infty$.

Evaluating the expression eq.(\ref{lapexa}) at $t=0$ we arrive at
the sum rule
\begin{equation}\label{sumrule}
\mathcal{Z}_k = 1- \left[\omega^2_k+\Sigma_k(0)\right] \;
\int_{-\infty}^{+\infty} \frac{d\omega}{\omega} \;
\frac{\rho_k(\omega) \; }{\left[
\omega^2-\omega^2_k- \mathrm{Re}{\Sigma}_k(\omega) \right]^2
+\left[ \mathrm{Im}{\Sigma}_k(\omega)\right]^2}\; ,
\end{equation}
\noindent which is easily confirmed to lowest order in the
coupling $\lambda$ by using the expression for $\Sigma_k$ given by
eq.(\ref{realpart}), the expansions (\ref{expansions}) and
neglecting $\Sigma_k^{(1)}$ in the denominator on the right hand
side of eq.(\ref{sumrule}).

 The integral over the cut in
eq.(\ref{lapexa}) is usually too complicated to be computed in
close form. However, we can evaluate its asymptotic behavior for
late times provided we know the behavior of $\rho(k,\omega)$ near
threshold.

For example, in the zero temperature case the integral over the
cut in eq.(\ref{lapexa}) extends over $ |\omega| > \omega_{th} $
where $\omega_{th} $ stands for the first production threshold.
Typical production cuts vanish at threshold with a power law at
zero temperature,
\begin{equation}\label{thresh}
 \rho_k(\omega) \buildrel{\omega \to \omega_{th}}\over = A \; \left(
 \omega - \omega_{th} \right)^{\alpha} \; ,
\end{equation}
where $ A $ is a positive constant. For example for the production
cut of two massive bosonic particles, the threshold behavior is
determined by $ \alpha = \frac12 $ and $ \omega_{th} = \sqrt{ (2
m)^2 + k^2} $.

We find from eq.(\ref{lapexa}) for late times,
\begin{eqnarray}\label{lapasi}
&&\delta({\vec k}, t)^{(cut)} \equiv
  \left[\omega^2_k+\Sigma_k(0)\right]  \ \;
\int_{-\infty}^{+\infty} \frac{d\omega}{\omega}
\frac{\rho_k(\omega) \; \cos(\omega t)}{\left[
\omega^2-\omega^2_k - \mathrm{Re}{\Sigma}_k(\omega) \right]^2 +
\left[\mathrm{Im}{\Sigma}_k(\omega)\right]^2} \buildrel{t \to
\infty}\over =  \; \frac{2\, A\;
\left[\omega^2_k+\Sigma_k(0)\right] }{\left[
\omega^2_{th}-\omega^2_k-\mathrm{Re} {\Sigma}_k(\omega_{th})
\right]^2 } \times \cr \cr &&\times \mbox{Re}
\int_{\omega_{th}}^{\infty} \frac{d\omega}{\omega}\; e^{i\omega t}
\left(\omega - \omega_{th} \right)^{\alpha} \buildrel{t \to
\infty}\over = \frac{2\, A\; (k^2 + m^2)\;
\Gamma(\alpha+1)}{\omega_{th} \; \left[  k^2 + m^2 - \omega_{th}^2
+ \mathrm{Re}\Sigma_k(\omega_{th}) \right]^2 } \,
\frac{1}{t^{\alpha+1}} \; \cos\left[ \omega_{th} t + \frac{\pi}{2}
(\alpha+1) \right]  \; .
\end{eqnarray}
We see that the cut contribution decreases for late times with a
power law. The smaller is the phase space near threshold (the
larger is $ \alpha $) the faster decreases the cut contribution
for late times\cite{nos1}.

Notice that eq.(\ref{lapasi}) holds provided the particle pole is
separated from the threshold. That is for $ \omega_{th} >
\Omega_k$.

\subsection{Resonances:}

In this case the imaginary part of the self energy does \emph{not}
vanish at $\omega_k$. The position of the (quasi) particle poles
is determined by the equation
\begin{equation}
s^2_p+\omega^2_k+\widetilde{\Sigma}_k(s_p)=0  \; .
\label{quasipole}
\end{equation}
Writing $s_p=-i \, \Omega_k + \gamma_k~;~ \gamma_k \ll \Omega_k$
with both $\Omega_k$ and $\gamma_k$ real, and comparing real and
imaginary parts we find
\begin{eqnarray}
&& \Omega^2_k -\omega^2_k -\mathrm{Re}\Sigma_{k} (\Omega_k) = 0
\quad , \quad \gamma_k = -\mathrm{sign}(\gamma_k) \; \pi \; 
\frac{\rho_k(\Omega_k)}{2\Omega_k}
\label{imagquasi}
\end{eqnarray}
The first equation is the same as in the previous case
eq.(\ref{polo}). Since,
$\frac{\rho(\Omega_k)}{\Omega_k}>0$ unless there
are instabilities,  it is clear that there is \emph{no solution}
for the second equation. Thus there is no complex pole in the
first (physical) Riemann sheet, corresponding to a resonance or
quasi-particle. Since there are no poles in the first (physical)
Riemann sheet in the complex $s$-plane, the Laplace transform is
obtained by wrapping the contour around the imaginary axis with
the result
\begin{equation}\label{laplareso}
\delta({\vec k}, t) \equiv \frac{{ \phi}_k( t)}{{
    \phi}_k(0) } =     \left[\omega^2_k+\Sigma_k(0)\right] \; 
    \int_{-\infty}^{+\infty} \frac{d\omega}{\omega} \;
\frac{\rho_k(\omega) \; \cos(\omega t)}{\left[
\omega^2-\omega^2_k- \mathrm{Re}{\Sigma}_k(\omega) \right]^2
+\left[ \mathrm{Im}{\Sigma}_k(\omega)\right]^2} \; .
\end{equation}
In weak coupling, the integrand of (\ref{laplareso}) features a
sharp peak near $\omega \simeq \Omega_k$, near this peak the
integrand in eq.(\ref{laplareso}) is of the  Breit-Wigner form,
\begin{equation}\label{BW}
\frac{\pi \; \rho_k(\omega) }{\left[ \omega^2-\omega^2_k-
\mathrm{Re}{\Sigma}_k(\omega) \right]^2 +\left[
\mathrm{Im}{\Sigma}_k(\omega)\right]^2} \buildrel{\omega \simeq
\Omega_k}\over= \mathcal{Z}_k \;
\frac{\Gamma_k}{(\omega-\Omega_k)^2+\Gamma^2_k} \; ,
\end{equation}
with $\mathcal{Z}_k$ given by eq.(\ref{poleres}) and
\begin{equation}\label{Gamma}
\Gamma_k \equiv \frac{\pi \; \rho_k(\Omega_k)}{2\Omega_k
\left[1+\frac{1}{2\Omega_k}\frac{\partial \mathrm{Re}
{\Sigma}_k(\omega)}{\partial \omega}\Big|_{\omega=\Omega_k}
\right]} \; .
\end{equation}
To lowest order in  weak coupling, $\Omega_k$ is given by
eq.(\ref{lopole}) and
\begin{equation}\label{ancho}
\Gamma_k=  \frac{\pi \; \rho_k(\omega_k)}{2 \; \omega_k }
+ {\cal O} (\lambda^2)  \; .
\end{equation}
Since these poles are in the unphysical Riemann sheet there is no
pole contribution deforming the contour $\Gamma$ in
eq.(\ref{invlap}). However, for small coupling the  poles in the
second Riemann sheet are near the cut and the sharp peak of the
resonance dominates the  cut contribution.

In such a case,  from eq.(\ref{invlap}) we find
\begin{equation}\label{lapins}
\phi_k(t)=  \mathcal{Z}_k \; \phi_k(0) \; e^{-\Gamma_k \, t} \;
\cos(\Omega_k t )+\mathcal{O}\Big(\frac{1}{t};\lambda^2 \Big) \; ,
\end{equation}
\noindent instead of eq.(\ref{lapexa}). Eq.(\ref{ancho}) exhibits
the Lorentz contraction of the width (dilation of lifetime) with
respect to that in the rest frame $(k=0)$.

\subsection{Threshold singularities and Bloch-Nordsieck
resummation}\label{sec:BN}

Let us consider the case in which the spectral density has a
threshold precisely at $\omega=\omega_k$, vanishing linearly at
threshold. A concrete realization corresponds to a  particle of
mass $m \neq 0$ that emits or absorbs a massless particle (at zero
temperature) with Lagrangian, \be\label{modelo} L = \frac12 \;
(\partial \Phi)^2 + \frac12 \; (\partial \pi)^2  - \frac12 \;
M^2\;  \Phi^2 - \frac{g}2 \; \Phi^2  \; \pi \;  \ee

\noindent which is a particular case of the specific examples
analyzed in section (\ref{sec:specexa}).

The self-energy to one-loop takes here at zero temperature the
form,
\begin{equation}\label{selfenethre}
\tilde{\Sigma}_k(s)= \frac{g^2}{(4\, \pi)^2}\left(1 +
\frac{m^2}{s^2+k^2}\right) \log\left(1 +\frac{s^2+k^2}{m^2}\right)
+ \mathcal{O}\Big(g^4 \Big) = \lambda \; m^2 \;
\frac{s^2+k^2+m^2}{s^2+k^2} \;
 \log\left(\frac{s^2+k^2+m^2}{m^2} \right) + \mathcal{O}\Big(\lambda^2
 \Big)\; .
\end{equation}
where we identify the dimensionless coupling $ \lambda \equiv
\frac{g^2}{(2 \, \pi \, m)^2} $, and have adjusted the mass
counterterm to cancel an ultraviolet logarithmic divergence.

 The wave function renormalization is UV finite
but features a logarithmic \emph{infrared} divergence at
$s^2=-(k^2+m^2)$. That is, $\frac{\partial
  \tilde{\Sigma}_k(s)}{\partial s}$ is divergent at $s^2=-(k^2+m^2)$.
This infrared divergence entails that there is no longer an
isolated particle pole, which now becomes the beginning of a
logarithmic branch cut as a consequence of the emission and
absorption of massless quanta.

The spectral density to lowest order in $\lambda$ is given by
\begin{equation}\label{specthresh}
\rho^{(1)}_k(\omega)= m^2 \; \frac{\omega^2-k^2-m^2}{\omega^2-k^2}
\; \Theta(\omega^2-k^2-m^2)~\mathrm{sign}(\omega) \; .
\end{equation}
 We find that the inverse propagator in the Laplace
transform in eq.(\ref{sols})  is given by
\begin{equation}\label{invprop}
s^2+k^2+m^2+\tilde{\Sigma}_k(s)=  (s^2+k^2+m^2)\left[1+\lambda \;
  \frac{m^2}{s^2+k^2}\log\left(\frac{s^2+k^2+m^2}{m^2}  \right)\right]
  +\mathcal{O}\Big(\lambda^2 \Big)\; .
\end{equation}
The form of the inverse propagator given by (\ref{invprop})
indicates  that in this case the logarithmic threshold behavior is
associated with a wave function renormalization.

 This situation
is common to a wide variety of theories, in particular in Quantum
Electrodynamics where the absorption and emission of massless
photons introduces logarithmic threshold divergences in the
electron propagator\cite{weinberg} which are enhanced at finite
temperature\cite{blaizot,drgqed}. The logarithmic infrared
divergence signals the breakdown of the perturbative expansion
near the particle mass shell. The Bloch-Nordsieck
method\cite{weinberg} provides a resummation of the leading
logarithmic contributions and is equivalent to a
\emph{renormalization group improvement of the propagator and the
self energy}. It leads\cite{weinberg} to the following resumed
form of the inverse propagator near the mass shell
$s^2+k^2+m^2\rightarrow 0$,
\begin{equation}\label{invpropBN}
s^2+k^2+m^2+\widetilde{\Sigma}_k(s)\rightarrow
(s^2+k^2+m^2)^{1-\lambda} \; .
\end{equation}
The inverse Laplace transform with this resummed propagator [see
eq.(\ref{invlap}] can be done  by deforming the contour leading to
the following integral representation,
\begin{equation}\label{bn}
\phi_k(t) = \phi_k(0) \; \frac{2 \; \sin \pi\lambda }{\pi} \;
\int_{\omega_k}^\infty \frac{d\omega}{\omega}\frac{\cos\omega
t}{\left(\frac{\omega^2}{\omega^2_k}
  - 1\right)^{1-\lambda}}
\end{equation}
While the integral can be computed exactly in terms of Bessel and
Struve functions\cite{drg1}, the asymptotic long time limit is
dominated by the threshold behavior at the lower bound of
integration. We find for $mt \gg 1$,
\begin{equation}\label{asythre}
\phi_k(t) \buildrel{mt \gg 1}\over = \phi_k(0) \;
\left(\frac{2}{\omega_k t} \right)^{\lambda} \;
\frac{\cos\left(\omega_k t
  + \frac{\pi \, \lambda}{2} \right)}{\Gamma(1-\lambda)}\left[ 1
  +\mathcal{O}\Big(\frac{1}{\omega_k \; t} \Big) \right] \; .
\end{equation}
Thus a Bloch-Nordsieck or renormalization group resummation of
threshold infrared divergences leads to an asymptotic long time
relaxation featuring an \emph{anomalous} dimension.

A perturbative expansion of eq.(\ref{asythre}) in powers of the
coupling $\lambda$ leads to
\begin{equation}\label{pertsec}
\phi_k(t)= \phi_k(0)\cos(\omega_kt)\left[1-\lambda \; \ln
\omega_kt +\mathcal{O}\Big(\lambda^2,\frac{1}{\omega_k \;
  t}\Big)\right]  \; ,
\end{equation}
\noindent the logarithmic \emph{secular} term signals the
breakdown of the perturbative expansion. The Bloch-Nordsieck
resummation of the infrared threshold divergences in the
propagator leads to relaxation in terms of a power law with an
anomalous dimension $\lambda$. We will see below that the
logarithmic secular term in eq.(\ref{pertsec}) will emerge in the
perturbative expansion in real time and that the dynamical
renormalization group will provide a resummation of these secular
terms leading to the same result as obtained from the
Bloch-Nordsieck resummation, given by eq.(\ref{asythre}) above.

Anomalous power law relaxation has  been found in condensed matter
physics in models of mesoscopic transport\cite{schon} and in the
X-ray absorption edge singularities in metals\cite{mahan} where
the spectral density also features logarithmic singularities at
threshold\cite{mahan}, as well as in the relaxation of fermionic
excitations in cold dense QCD\cite{boyadense}.

\subsubsection{Threshold divergences}

Consider the case in which the spectral density near $\omega =
\omega_k$ is of the form
\begin{eqnarray} \rho^{(1)}_k(\omega) & = &
m^2 \; \left[\frac{\omega^2-k^2-m^2}{\omega^2-k^2}+ b_k
\right]\Theta(\omega-\omega_k) \; \mbox{sign}(\omega) \; ,
\label{Tspecdens}
\end{eqnarray}
\noindent with $b_k\neq 0$. This corresponds to the nonzero
temperature case in the model (\ref{modelo}).

 Following the steps outlined above, we now find that near the
 particle mass shell the inverse propagator is now given by,
\begin{equation}\label{invpropT}
s^2+k^2+m^2+\widetilde{\Sigma}_k(s) \buildrel{\omega \to
\omega_k}\over= (s^2+k^2+m^2)\left[1+\lambda \;
\frac{m^2}{s^2+k^2} \ln\left(\frac{s^2+k^2+m^2}{m^2}
\right)\right]-\lambda \; m^2 \; b_k \;
\ln\left[\frac{s^2+k^2+m^2}{m^2} \right]+\mathcal{O}(\lambda^2)
\end{equation}
\noindent and we have  chosen $Z=1$. Whereas the logarithmic
divergence near the particle mass shell (from the zero temperature
contribution) can be interpreted in terms of a wave function
renormalization and exponentiated via the Bloch-Nordsieck or
renormalization group resummation, the nonzero temperature
contribution  $b_k$ leads to a stronger divergence, since not only
the derivative of the inverse propagator but now the  inverse
propagator itself diverges at threshold. This strong divergence
cannot be resummed either by the renormalization group or
Bloch-Nordsieck resummation.

As we shall see below, the dynamical renormalization group in real
time provides in this case a resummation  that extracts the
asymptotic long time behavior.

\section{Dynamical renormalization group:}\label{section:DRG}

We obtain here the perturbative expansion of the field expectation
value and its late time behavior exhibiting secular terms. Then,
we resum the secular terms using the DRG.

\subsection{Perturbative Expansion and Secular Terms}

Expanding the Laplace transform eq.(\ref{sols}) in powers of $
\lambda $ yields,
\begin{equation}
\varphi_k(s)  = \phi_k(0) \; \Bigg\{ \frac{s}{s^2+\omega^2_k} +
\lambda
  \; \left[ \frac{\widetilde{\Sigma}^{(1)}_k(s) -
      \widetilde{\Sigma}^{(1)}_k(0)}{s \; 
  (s^2+\omega^2_k)} - \frac{ s \;
  \widetilde{\Sigma}^{(1)}_k(s)}{(s^2+\omega^2_k)^2} \right] +
  \mathcal{O}(\lambda^2)\Bigg\} \; .
\label{etaexp}
\end{equation}
The real time evolution follows by inverse Laplace transform order
by order in perturbation theory. It is convenient first to express
the self-energy as a dispersion relation [see
  eq.(\ref{laplasigma})] yielding,
\be \label{lap1} 
\varphi_k(s)  = \phi_k(0) \; \Bigg\{
\frac{s}{s^2+\omega^2_k} - \lambda
  \; \left[ \frac{s \; \widetilde{\Sigma}^{(1)}_k(0)}{(s^2+\omega^2_k)^2} 
- 2 \; \omega^2_k \; 
\int_0^{\infty}\frac{d\omega}{\omega} \; \rho^{(1)}_k(\omega) \; \frac{
  s}{(s^2+\omega^2_k)^2 \, (s^2+\omega^2)} \right] +
  \mathcal{O}(\lambda^2)\Bigg\} \; .
\ee 
The inverse Laplace transform of eq.(\ref{lap1}) is easily
obtained comparing with the elementary integrals, 
\bea\label{integ}
&&\frac{s}{s^2+\omega^2_k} = \int_0^{\infty} dt \; e^{-st} \;
\cos\omega_k t \; , \cr \cr 
&&\frac{s}{(s^2+\omega^2_k)^2} 
= \frac{1}{2 \; \omega_k}
\int_0^{\infty} t \; dt \; e^{-st} \;  \sin\omega_k  t \; ,  \cr \cr 
&&\frac{s}{(s^2+\omega^2_k)^2 \,   (s^2+\omega^2)} =
\frac{1}{(\omega^2 - \omega_k^2)^2} \int_0^{\infty} dt \; e^{-st}
\; [\cos\omega t - \cos\omega_k
  t + \frac{\omega^2 - \omega_k^2}{2 \, \omega_k} \, t \;
  \sin\omega_k t ]\; .
\eea We find comparing eqs.(\ref{lap1} ) and (\ref{integ}) and then using
eq.(\ref{laplasigma}),  
\be 
\phi_k(t) = \phi_k(0) \; \Bigg\{ \cos \omega_k t  + \lambda
  \; \left[2 \, \omega^2_k \; \mathcal{P}\int_0^{\infty}
    \frac{d\omega}{\omega} \;
\frac{\rho^{(1)}_k(\omega)}{(\omega^2 - \omega_k^2)^2} (\cos\omega
t - \cos\omega_k   t) - \frac{t}{2 \; \omega_k} \; \sin\omega_k t \;
\mathrm{Re}{\Sigma}^{(1)}_k(\omega_k) \right] +
  \mathcal{O}(\lambda^2)\Bigg\} \; .
\ee 
The second term of order $\lambda$ turns to be the frequency shift $
\delta \omega_k $ given by eq.(\ref{lopole}). 

To first order in $\lambda$ we finally have, 
\be \label{general} 
\phi_k(t) = \phi_k(0) \; \Bigg\{ \cos \omega_k
t  - t \; \delta \omega_k \; \sin\omega_k t + 2 \, \lambda
  \;  \omega^2_k \; \mathcal{P}\int_0^{\infty} \frac{d\omega}{\omega} \;
\frac{\rho^{(1)}_k(\omega)}{(\omega^2 - \omega_k^2)^2}\;
(\cos\omega t - \cos\omega_k   t) +  \mathcal{O}(\lambda^2)\Bigg\}
\; . \ee The second term is a secular term: it grows as $t$.
Furthermore, the frequency integral  in the third term may also
feature secular terms  at long times. Its $t$ dependence for late
times is determined by the behavior of $\rho^{(1)}_k(\omega)$ for
$ \omega \to \omega_k$ as we analyze below in several relevant
cases.

 \subsection{Dynamical renormalization group}

In perturbation theory we find that the solution of
 the evolution equation has in general the asymptotic form,
 \begin{equation}\label{secu}
 \phi_k(t)=\frac{1}{2} \; \phi_k(0) \; e^{i\omega_k t}\Bigg[1+\lambda \;
 \mathcal{S}^{(1)}_k(t)+\lambda^2 \;
 \mathcal{S}^{(2)}_k(t) +\mathrm{non~
 secular}+\mathcal{O}(\lambda^3)\Bigg]+ \mathrm{c.c.}
 \end{equation}
 \noindent with $\mathcal{S}^{(1,2)}_k(t)$ containing only secular
 terms, namely terms that grow in time.

The dynamical renormalization group absorbs the secular terms into
a \emph{renormalization} of the complex amplitude at an arbitrary
time scale $\tau$. This is achieved by introducing a
multiplicative renormalization  in the following manner,
\begin{equation}
\phi_k(0)= \phi_k(\tau) \; R_k(\tau)\label{renoconstant}
\end{equation}
\noindent with $R_k(\tau)$ found systematically in an expansion in
the coupling $\lambda$, namely
\begin{equation}
R_k(\tau) = 1+ \lambda \; r^{(1)}_k(\tau)+ \lambda^2 \;
r^{(2)}_k(\tau)+{\cal O}(\lambda^3) \label{Rexpa}
\end{equation}
The coefficients $r^{(i)}_k(\tau)$, which play the r\^ole of
 counterterms, are chosen to cancel the secular
 terms  at the time $t=\tau$ order by order in
 the perturbative expansion. To lowest order in $\lambda$ we find,
 \begin{equation}
r^{(1)}_k(\tau)=-\mathcal{S}^{(1)}_k(\tau)
 \end{equation}
 The solution renormalized in this way at a scale $\tau$ is to lowest order in
 $\lambda$ given by
 \begin{equation}\label{fi1}
\phi_k(t)=\phi_k(\tau)\left\{1+\lambda\left[
\mathcal{S}^{(1)}_k(t)-\mathcal{S}^{(1)}_k(\tau)\right]+
\mathrm{non~secular}+\mathcal{O}(\lambda^2) \right\}+\mathrm{c.c.}
 \end{equation}
Since $\tau$ is an arbitrary scale, the solution must not depend
on it. We therefore request
 \begin{equation}\label{DRG}
 \frac{\partial}{\partial \tau} \phi_k(t)=0 \; .
 \end{equation}
This is the dynamical renormalization group equation.

Inserting eq.(\ref{fi1}) into eq.(\ref{DRG}) to lowest order in
the coupling, the dynamical renormalization group equation
becomes,
 \begin{equation}
\frac{\partial }{\partial \tau}
\ln[\phi_k(\tau)]=\mathcal{S}^{(1)}_k(\tau) \; .
\end{equation}
\noindent with solution
\begin{equation}
\phi_k(\tau)=\phi_k(\tau_0)\, e^{\mathcal{S}^{(1)}_k(\tau)-
\mathcal{S}^{(1)}_k(\tau_0)}+\mathrm{c.c.}\; .
\end{equation}
Since $ \phi_k(t) $ is independent of $\tau$ thanks to the DRG
eq.(\ref{DRG}), we can choose the arbitrary scale $\tau$ to
coincide with $t$. This yields,
\begin{equation}
\phi_k(t) \buildrel{t \to  \infty}\over= \phi_k(0) \,
e^{\mathcal{S}^{(1)}_k(t)}\left[1+ \mathcal{O}\left(\lambda^2,
 \frac{1}{t}\right) \right]
+\mathrm{c.c.} \label{finDRGsol}
\end{equation}
This is the long time asymptotic behavior given by \emph{dynamical
renormalization group}. The overall multiplicative normalization
can be adjusted up to $ \mathcal{O}\left(\lambda^2\right) $ using
eq.(\ref{secu}). The next sections extract the leading secular
term $\mathcal{S}^{(1)}_k(t)$ in several cases.

\subsection{Case I: Particles and quasiparticles}

 We consider first the case where the frequency $\omega_k$ is away
 from the multiparticle thresholds, namely
 \begin{equation}
 \omega_k \neq \omega_{th}~~;~~ \rho_k(\omega>0) \neq 0
 ~~\mathrm{for}~~\omega_{th}<\omega <+\infty \; .
 \end{equation}
That is, the spectral density is continuous, finite and non-zero
at $ \omega = \omega_k$.
 This case treats on equal footing the situations corresponding to
 a stable particle $\omega_k < \omega_{th}$ and that of a
 quasiparticle $\omega_k > \omega_{th}$.

 The long time limit $t\rightarrow \infty$ can be extracted from
 eq.(\ref{general}) by  using formulae (\ref{formula1}) in appendix
 \ref{appendix:formulae}. We find the asymptotic long time
 behavior to be given in this case by
 \begin{eqnarray}
\phi_k(t) & = & \frac{\phi_k(0)}{2} \; e^{i\omega_kt} \; \Bigg\{ 1
+ \lambda \; \Bigg[ t \; \frac{i
\,\mathrm{Re}\Sigma^{(1)}_k(\omega_k)}{2\omega_k}- t \;
\frac{\pi \; \rho^{(1)}_k(\omega_k)}{2\omega_k}+ \nonumber
\\ & - & \mathcal{P} \int d\omega \;
\frac{\rho^{(1)}_k(\omega)}{\omega}\frac{\omega^2_k}{(\omega^2
-\omega^2_k)^2}+\mathcal{O}\left(\frac{1}{t}\right) \Bigg] +
\mathcal{O}(\lambda^2) \Bigg\} +c.c. \label{secular1} \; .
 \end{eqnarray}
The perturbative solution displays \emph{secular terms} that in
this case grow linearly in time. The secular terms invalidate the
perturbative expansion for late times $ t \gg \frac{1}{m
\;\lambda}$.

Clearly, a resummation scheme must be invoked to resum the
perturbative expansion.

In the notation of eq.(\ref{secu}) we have, \be
\mathcal{S}^{(1)}_k(t) =  \frac{t}{2\omega_k} \; \left[ i
\,\mathrm{Re}
\Sigma^{(1)}_k(\omega_k)-\pi \; \rho^{(1)}_k(\omega_k)
\right]\; . \ee Therefore, eq.(\ref{finDRGsol}) for the DRG
solution in the long time limit takes here the form,
\begin{eqnarray}
\phi_k(t) & = &  \mathcal{Z}_k \;
\phi_k(0)~\cos([\omega_k+\delta\omega_k]t)
 \; e^{-\Gamma_k \;t} + \mathcal{O}\left(\frac{1}{t},\lambda^2 \right) \; ,
 \label{solren1} \cr \cr
\delta\omega_k & = &\lambda \;
\frac{\mathrm{Re}\Sigma^{(1)}_k(\omega_k)}{2\omega_k}~~;~~\Gamma_k
=\lambda \; \frac{\pi \; \rho^{(1)}_k(\omega_k)}{2\omega_k}\label{masswidth}
\; ,
\end{eqnarray}
This can also be written as,
\begin{equation}\label{RGimproved1}
\phi_k(t)  =  \mathcal{Z}_k \; \phi_k(0)~\cos(\Omega_k \,  t) \;
e^{-\Gamma_k \; t} + \mathcal{O}\left(\frac{1}{t},\lambda^2\right)
\; ,
\end{equation}
\noindent with $\Omega_k, \; \mathcal{Z}_k$ and $\Gamma_k$ given
by equations (\ref{lopole}), (\ref{Zlo}) and (\ref{ancho})
respectively.

Comparing this DRG solution with that obtained from the Laplace
transform, given by eq.(\ref{lapins}), we see that in this case
the dynamical renormalization group provides a resummation akin to
that of the Dyson series.

If $\rho_k(\omega_k)=0$, namely the (stable) particle pole is
below the multiparticle threshold, then $\Gamma_k=0$ and the
solution describes the asymptotic time evolution of a stable
particle. On the other hand if $\omega_k$ is above the
multiparticle threshold, then the pole moves off the physical
sheet into the second Riemann sheet becoming a quasiparticle pole.
This case describes a decaying particle or resonance.

\subsection{Threshold singularities:}

We consider in this subsection the case where the spectral density
is discontinuous at $\omega = \omega_k$ (case II) or it exhibits a
logarithmic singularity at this point (case III).

\subsubsection{Case II: a discontinuous density at $\omega = \omega_k$.}

 Let us consider the case in which the spectral density around
 $\omega = \omega_k$ is given by equation (\ref{Tspecdens}).

This case reflects a \emph{discontinuity} at threshold since
$\rho^{(1)}_k(\omega=\omega_k-0)=0;\rho^{(1)}_k(\omega=\omega_k+0)=A_k
\; b_k$ and near threshold $\omega \sim \omega_k$ the spectral
density is approximated by
\begin{equation}
\rho^{(1)}_k(\omega)= A_k\left[\omega-\omega_k+b_k\right]\left[ 1
+
  {\cal O}(\omega-\omega_k) \right] \;
\Theta(\omega-\omega_k)~~,~~ A_k, \, b_k >0 \label{nT}\; .
\end{equation}
We can now extract the asymptotic long time behavior for $\phi(t)$
given by equation (\ref{general}) by using the results given by
eqs. (\ref{formula2}), (\ref{formula6}) and (\ref{formula7}) in
appendix B. We find, \bea\label{fiumb} &&\phi_k(t) \buildrel{t \to
\infty}\over =  \frac{\phi_k(0)}{2} \; e^{i\omega_kt} \left\{ 1 +
i \; t \left[ \delta \omega_k - \lambda \; \frac{A_k \;
b_k}{2\omega_k} \log(\omega_k \; t \; e^{\gamma -1}) \right]
\right. \cr \cr &&\left. - \lambda \; \frac{A_k}{2\omega_k}\left[
\frac{\pi}{2} \; t \;
  b_k + \left(1
- \frac{2 \; b_k}{\omega_k} \right) \log(\omega_k \; t \;
e^{\gamma -1}) \right] + \mathcal{O}(t^0)  \right\} +c.c +
\mathcal{O}(\lambda^2) \; , \eea With the convention
$\Theta(0)\equiv \frac{1}{2}$ we see that the real linear secular
term is given by $\Gamma^{(1)}_k \, t \; $ with $
\Gamma^{(1)}_k=-\mathrm{Im}\Sigma^{(1)}_k(\omega_k)/2\omega_k
=\frac{\pi \; A_k \; b_k}{4 \; \omega_k} $.

In particular, for $b_k=0$ we identify the secular term $-\lambda
\; \ln[\omega_k \, t]$ which is manifest in the
\emph{perturbative} expansion eq.(\ref{pertsec}) of the
Bloch-Nordsieck resummed result given by eq.(\ref{asythre}).

Up to $\mathcal{O}(\lambda)$ we now find  eq.(\ref{fiumb}) given
by eq.(\ref{secu}) with the secular term,
\begin{equation}\label{secT}
\mathcal{S}^{(1)}_k(t)= i \; t \left[ \delta \omega_k - \lambda \,
\frac{A_k \;
    b_k}{2 \, \omega_k} \log(\omega_k \; t \; e^{\gamma -1}) \right]
-\lambda \; \frac{A_k}{2 \, \omega_k}\left[ \frac{\pi}{2} \; t \;
b_k +\left(1 -\frac{2 \; b_k}{\omega_k} \right) \, \log(\omega_k
\; t \; e^{\gamma
    -1}) \right] \; .
 \end{equation}
The DRG resummation eq.(\ref{finDRGsol}) takes here the form,
\begin{equation}\label{asyT}
\phi_k(t) \buildrel{t \to \infty}\over = \phi_k(0) \;
\cos[\Xi_k(t) \; t] \; e^{-\Gamma_k t} \; \left[\frac{t_0}{ t}
  \right]^{\nu_k} + \mathcal{O}\Bigg(\frac{1}{t};\lambda^2\Bigg) \; .
\end{equation}
\noindent $\Xi_k(t), \; \Gamma_k$ and $\nu_k$ are given by
\begin{eqnarray}
\Xi_k(t) & = & \omega_k+ \delta \omega_k + \frac{\lambda \, A_k \,
b_k}{2 \, \omega_k} \ln\left(\omega_k \; t\; e^{\gamma -1}\right)
\label{phase} \; ,\cr \cr \nu_k & = & \frac{\lambda  \, A_k}{2 \,
\omega_k}\left( 1 - \frac{2 \,
  b_k}{\omega_k}\right) ~~,~~  \Gamma_k  =
\frac{\pi \,  \lambda \,  A_k\,b_k}{4 \, \omega_k} \;
.\label{widthT}
\end{eqnarray}
Thus, we see that the logarithmic mass shell singularity
proportional to $b_k$ in the perturbative expansion of the
propagator, eq.(\ref{invpropT}), is manifest in the real-time
perturbative expansion given by eq.(\ref{secT}) as a secular term
of the form $t \ln[t]$. The dynamical renormalization group resums
this secular term into a logarithmic phase in the field
expectation value eq.(\ref{asyT}).

Notice from eqs.(\ref{nT}) and (\ref{specthresh}) that $ A_k = 2
\, \omega_k$ and $b_k=0$ for the model (\ref{modelo}). Hence, we
see in eq.(\ref{widthT}) that $ \nu_k = \lambda $ in this case.
Therefore, eqs.(\ref{asythre}) and (\ref{asyT}) fully agree. In
addition, eq.(\ref{asythre}) implies $ t_0 = 2 \,
e^{-\gamma}/\omega_k $ for the reference time scale.

Thus, an important conclusion resulting from the comparison
between the time evolution obtained from the  Bloch-Nordsieck
resummed propagator given by eq.(\ref{asythre}) for $b_k=0$  and
the dynamical renormalization group resummation in this case, is
that the DRG resummation is a Bloch-Nordsieck resummation \emph{in
real time}.

Moreover, while a Bloch-Nordsieck resummation of the Euclidean
propagator for $b_k \neq 0$ given by eq. (\ref{invpropT}) is not
available, the dynamical renormalization group is able to resum
the case $b_k\neq 0$.

\subsubsection{Case III: a logarithmically divergent spectral density at $\omega
  = \omega_k$.}

We now consider the case in which the spectral density is
logarithmically divergent at $\omega_k$ but does \emph{not}
feature a normal threshold at this value of frequency, namely
\begin{equation}
\rho^{(1)}_k(\omega) \buildrel{\omega \to \omega_k}\over = C_k \;
\left[ \log\left|\frac{\mu}{\omega-\omega_k} \right| +{\cal
O}(\omega-\omega_k) \right] \;~~;~~ C_k>0 \; . \label{logaspec}
\end{equation}
In this case, using formulae (\ref{formula3})-(\ref{formula4})
from appendix \ref{appendix:formulae}, we find that the field
expectation value takes the form of eq.(\ref{secu}) where the
secular term $\mathcal{S}^{(1)}_k(t)$ is given by
\begin{equation}\label{singlog} \lambda \;
\mathcal{S}^{(1)}_k(t)=  i\left[  t \; \delta \omega_k
-\frac{\lambda \, \pi \; C_k}{\omega_k^2} \; \ln\left(\mu \;
t~e^{\gamma-1}\right)\right] -\frac{\lambda \; \pi \;  C_k}{2 \,
  \omega_k} \; t \;  \ln\left[\mu \;  t \, e^{\gamma-1}\right]
\end{equation}
Inserting eq.(\ref{singlog}) in eq.(\ref{finDRGsol}) we find the
DRG asymptotic long time behavior,
\begin{equation}
\phi_k(t) \buildrel{t \to \infty}\over= \phi_k(0) \;
\cos[\Omega_k(t)]
 \; e^{-\gamma_k \;  t  \; \ln[\overline{\mu}t]} + \mathrm{c.c.} \; ,
\end{equation}
\noindent with
\begin{equation}
\Omega_k(t) = (\omega_k+\delta \omega_k)t -\frac{\lambda \, \pi \;
C_k}{\omega_k^2} \; \ln(\overline{\mu} t)
\end{equation}
\noindent and
\begin{equation}
\gamma_k =\frac{\pi \; \lambda \;  C_k}{2 \; \omega_k}~~,~~
\overline{\mu}= \mu \; e^{\gamma-1} \;  , \;  \delta \omega_k  =
\frac{\lambda}{2\omega_k} \; \mathrm{Re}{\Sigma}^{(1)}_k(\omega_k)
\; .
\end{equation}

 \subsection{Summary of dynamical renormalization group
 resummation to lowest order}

We are now in position to summarize the main results obtained to
lowest order in the dynamical renormalization group resummation
program.

The nature of the secular terms is completely determined by the
behavior of the spectral density $\rho_k(\omega)$  near
$\omega=\omega_k$. The different cases studied above to lowest
order in the coupling $\lambda$ describe fairly general behavior
of the spectral densities in quantum field theories at zero and
non-zero temperature.

The important aspects of the results of this study can be
highlighted as follows

\begin{itemize}
\item{ {\bf i):} when the spectral density $\rho_k(\omega)$
vanishes at $\omega=\omega_k$ the particle is stable and the only
effect of the interaction is to renormalize the mass and wave
function. This corresponds to the spectral density of the particle
featuring an isolated pole away from  the branch cuts associated
with the multiparticle production. The resummation implied by the
dynamical renormalization group is equivalent to the Dyson
resummation of the propagator that leads to the renormalized
dispersion relation for the particle. When
$\rho_k(\omega=\omega_k)\neq 0$ but $\omega_k
> \omega_{th}$ with $\omega_{th}$ being the multiparticle thresholds,
namely the beginning and end points of branch cuts in the spectral
density of the particle, the particle is unstable and its
propagator features a pole in an unphysical sheet in the complex
$\omega$ plane, namely a resonance. In perturbation theory this
resonance is very close to the real axis and the propagator is of
the Breit-Wigner form. The dynamical renormalization group in this
case provides a resummation which is equivalent to the Dyson
series. }

\item{ {\bf ii):} When the position of the multiparticle
thresholds coincide with $\omega_k$, infrared divergences arise
near the particle mass-shell. This  corresponds to the emission
and absorption of massless particles, a situation which is common
in gauge theories and in theories with spontaneous symmetry
breaking with Goldstone bosons. The secular terms that determine
the relaxation of the mean field in perturbation theory depend on
the behavior of the spectral density near thresholds. The
dynamical renormalization group resummation in the case
$\rho_k(\omega_k \pm 0 )=0$ is similar to the Bloch-Nordsieck
resummation of infrared divergences but directly in \emph{real
time}. No Bloch-Nordsieck resummation is available for the case
II, which we solve by DRG and where the spectral density is
discontinuous ($\rho_k(\omega_k + 0 ) \neq 0 $). Furthermore, the
DRG provides a consistent resumation in the case III with a
\emph{logarithmically divergent} spectral density at
$\omega=\omega_k$.}

\end{itemize}

The secular terms associated with the different behaviors of the
spectral density for these generic cases are summarized in the
table below.


\begin{turn}{90}
\begin{tabular}{|c|c|c|c|c|}\hline
   & &  & DRG & Resummation \\
Case  &\hspace{0.3cm} $\rho_k(\omega) \buildrel{\omega \to
\omega_k}\over =$&\hspace{1.0cm} Secular term  $
\mathcal{S}^{(1)}_k(t) $ &resummed amplitude &\hspace{0.4cm}
done \\
 &   &  &$\delta({\vec k}, t)$ & by the DRG\\ \hline
 &  &   & &\\
I & $  \rho_k(\omega=\omega_k) \neq 0 ~;~ \rho_k(\omega=\omega_k
\pm 0)\neq 0$ &\hspace{0.9cm} $ \frac{t}{2 \, \omega_k }\left[i\,
\mathrm{Re}\Sigma^{(1)}_k(\omega_k)+\mathrm{Im}\Sigma^{(1)}_k(\omega_k)\right]$
&   $ e^{it\left( \omega_k + \delta \omega_k\right)- t \,
\frac{\pi \, \rho_k(\omega_k)}{2 \, \omega_k}} $ & Dyson
\\ & and continuous, $\omega_k\neq \omega_{th} $  &   & & \\ &  & &  &
\\ \hline   &  &   & &\\
II & $A_k[\omega-\omega_k+b_k]\Theta(\omega-\omega_k)$ &
$\frac{i\,t}{2 \, \omega_k
}\left[\mathrm{Re}\Sigma^{(1)}_k(\omega_k)-A_k\,b_k\ln\left(\omega_k
\; t \; e^{\gamma-1}\right)\right]$ &  $e^{it\left[ \omega_k +
\delta \omega_k -\lambda \, \frac{A_k\,b_k}{2 \, \omega_k} \;
\ln\left(\omega_k\; t \; e^{\gamma-1}\right)\right]}$
& Dyson $ + $ \\
 &  $A_k , \; b_k>0 $   & $-\frac{A_k}{2  \;\omega_k} \left(1
  - \frac{2 \; b_k}{\omega_k} \right) \ln\left(\omega_k \;
t\;e^{\gamma-1}\right) -t \;
\frac{\pi \, \rho^{(1)}_k(\omega_k)}{4\omega_k} $ & $ \times
\; e^{- t \, \frac{\pi \,
\rho_k(\omega_k)}{2 \, \omega_k}} $&  Bloch-Nordsieck    \\
 &  &   & & for the kernel\\ \hline
 &  &   & &\\
III & $C_k \ln\left|\frac{\mu}{\omega - \omega_k}\right|$ &
$\frac{i}{2 \, \omega_k }\left[ t \;
\mathrm{Re}{\Sigma}_k(\omega_k)-\frac{2\pi
C_k}{\omega_k}\ln\left({\bar\mu}\,t\right)\right] $ &  $
e^{it\left( \omega_k + \delta \omega_k\right)- \frac{\pi\,
    C_k}{2 \, \omega_k} \, t \,\log\left({\bar\mu}\,t\right)} \;  $
& Dyson $  + $ \\
& $C_k>0$  & $-\frac{\pi C_k}{2 \, \omega_k} \,t \ln({\bar\mu}\, t
)$ &  $\times \; \left({\bar\mu}\,t\right)^{-\frac{i \, \pi
C_k}{\omega_k^2}}$
& Bloch-Nordsieck \\ &  &   & & for the kernel\\
\hline
\end{tabular}
\end{turn}

\bigskip

{TABLE 1. Secular terms to first order in perturbation theory for
the four cases considered and their DRG resummation. The
convention $\Theta(0)=1/2$ is implied in case II above,
$\mathrm{Im}\Sigma_k(\omega_k)=-\pi \rho_k(\omega_k)$ and
${\bar\mu} = \mu \, e^{\gamma-1}$.}

\section{Specific examples in a scalar theory:}\label{sec:specexa}

To highlight the main concepts in a simpler setting we  study  a
scalar theory of three interacting fields $\Phi_i~;~i=1,2,3$. The
Lagrangian density is given by
\begin{equation}
{\cal L} = \sum_{i=1}^3\left[\frac{1}{2}(\partial_{\mu}\Phi_i)^2
-\frac{1}{2} \; m^2_{0,i} \;  \Phi^2_i \right]-g_o  \; \Phi_1 \;
\Phi_2 \; \Phi_3 +J_o \; \Phi_1 \; , \label{lagrascalar}
\end{equation}
\noindent where the source coupled to the  field $\Phi_1$ has been
introduced to provide an initial value problem for its expectation
value  as explained in section II, of course we could study
simultaneously the initial value problem for the expectation
values of all three fields, but our focus is to highlight the most
important aspects in a simpler setting.

The renormalization is performed  by introducing wave functions,
mass and coupling renormalizations as follows
\begin{eqnarray}
&&\Phi_i= Z^{1/2}_{i} \; \Phi_{i,r} \; ; \; J_r=Z^{1/2}_{1} \; J_o
\quad , \quad m^2_{0,i} \;  Z_{i} = m^2_{r,i} \;  Z_{i}+ \delta
m^2_{i} ~\; ; \;g_o \; \sqrt{Z_{1} \; Z_{2} \; Z_{3}}= g_r \;
Z_g\quad , \quad i=1,2,3. \label{renorma3field}
\end{eqnarray}
 We now suppress the
label $r$ with all quantities being renormalized, and write
 the Lagrangian density in terms of renormalized quantities
and counterterms,
\begin{eqnarray}
 {\cal L} & = & \sum_{i=1}^3\left[\frac{1}{2} \; (\partial_{\mu}\Phi_i)^2
 -\frac{1}{2} \; m^2_i \;  \Phi^2_i \right]-
 g \;  \Phi_1 \; \Phi_2 \; \Phi_3 +J \; \Phi_1+ {\cal L}_{ct} \nonumber \\
 {\cal L}_{ct} &  = &  \sum_{i=1}^3\left[\frac{1}{2} \; (Z_{i}-1) \;
(\partial_{\mu}\Phi_i)^2-m^2_i \; \Phi^2_i -\frac{1}{2} \; \delta
m^2_i \; \Phi^2_i \right] -g (Z_g-1) \; \Phi_1 \; \Phi_2 \; \Phi_3
\; .
\end{eqnarray}
The counterterms are adjusted in perturbation theory as usual.

Some cases which are relevant for the discussion that follows are:

\begin{itemize}
\item{{\bf i:}  with $m_1>m_2+m_3$, in this case the decay and
recombination processes $1\leftrightarrow 2+3$ are allowed. This
will lead to the presence of a resonance, namely the quasiparticle
pole will be in the continuum resulting in a decay rate. }

\item{{\bf ii:} $ m_1=m_2\neq 0\;; \, m_3=0$, this case will
feature anomalous logarithmic relaxation as a consequence of the
absorption and emission of soft massless quanta.}

\end{itemize}

In what follows, we will treat both cases by setting $m_3=0$ and
$m_1 = m$  to present the main results in a simpler and more clear
manner.

\begin{figure}[ht!]
\includegraphics[width=2.5in,keepaspectratio=true]{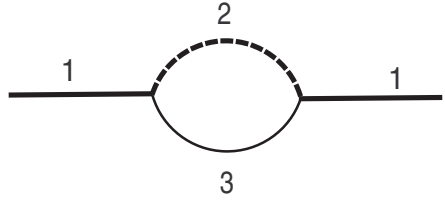}
\caption{One loop contribution to the self-energy of particle
$1$.}\label{fig:scalarloop}
\end{figure}

The calculation of the self-energy of particle $1$ at finite
temperature is best performed in the imaginary time formulation.
The Matsubara Green's functions for the fields are written as
spectral representations in the form
\begin{equation}\label{specrepG}
G_{i}(k,\omega_n) =
\frac{1}{\omega^2_n+\omega^2_{k,i}}=\int_{-\infty}^{+\infty} dp_0
\; \frac{\rho_i(p_0,k)}{p_0-i\omega_n}~~;~~\omega_n= {2\pi \,n \,
T}\; , \quad , \quad i=1,2,3.
\end{equation}
\noindent with the spectral densities
\begin{equation}
\rho_i(p_0,k)= \frac{1}{2\omega_k} \left[
\delta(p_0-\omega_{k,i})-\delta(p_0+\omega_{k,i})\right]~~,~~
\omega^2_{k,i}=k^2+m^2_i \quad , \quad
i=1,2,3\label{specdensparts}\; .
\end{equation}
The one-loop self-energy for the particle $1$ is shown in figure
\ref{fig:scalarloop} and  is given by
\begin{equation}\label{selfene1}
\Sigma_{1,k}(i\omega_n)= -g^2 \;  T \; \sum_{\omega_m}\int
\frac{d^3p}{(2\pi)^3} \;  dp_2 \;  dp_3 \;
\frac{\rho_2(p_2,p)}{p_2-i\omega_m} \;
\frac{\rho_3(p_3,|\vec{p}+\vec{k}|)}{p_3-i\omega_m-i\omega_n}  \;
.
\end{equation}
The sum over the Matsubara frequencies can be done using the
methods described in \cite{kapusta,lebellac} leading to the
following spectral representation  for the self-energy
\begin{equation}
\Sigma_{1,k}(i\omega_n) - \Sigma_{1,k}(0)
= i\omega_n \, \int_{-\infty}^{+\infty} d\omega' \;
\frac{\rho_k(\omega')}{\omega'(i\omega_n-\omega')}\; ,
\end{equation}
\noindent with the spectral density $\rho_k(\omega)$ given by
\begin{equation}
\rho_k(\omega)= g^2 \int \frac{d^3p}{(2\pi)^3} \;  dp_2\;  dp_3\;
\rho_2(p_2,p)\; \rho_3(p_3,|\vec{p}+\vec{k}|){\;
  [n(p_2)-n(p_3)]}\; \delta(p_3-p_2-\omega)~~;~~
n(p_i)=\frac{1}{e^{\frac{p_i}{T}}-1}\quad , \quad i=1,2,3 \; .
\end{equation}
Comparing this dispersive representation for the self-energy with
that of equation (\ref{disprel}) we find that
\begin{equation}
\Sigma_k(\omega)=\Sigma_{1,k}(i\omega_n=\omega+i0)
\end{equation}
A straightforward but lengthy calculation with $m_2=M \neq
0;m_3=0$ leads to the following spectral density
\begin{eqnarray}
\rho_k(\omega)=  && \lambda \; m^2 \,
\Bigg\{\left[\frac{\omega^2-W^2_k}{\omega^2-k^2}+\frac{T}{k} \log
\left( \frac{1-e^{-u^+}}{1-e^{-u^-}}
\frac{1-e^{-v^+}}{1-e^{-v^-}}\right)\right]\Theta(\omega^2-W^2_k)
+\nonumber \\ && +\frac{T}{k} \ln \left[
\frac{1-e^{-u^+}}{1-e^{u^-}}
\frac{1-e^{-v^+}}{1-e^{v^-}}\right]\Theta(k^2-\omega^2)\Bigg\} \;
\mathrm{sign}(\omega)\label{specdensFinT} \; ,
\end{eqnarray}
\noindent with
\begin{equation}
{u}^\pm = \frac{\omega^2-W^2_k}{2T( |\omega|\mp k)}~~,~~ {v}^\pm
=\frac{|\omega|}{T} - {u}^\mp \quad , \quad W^2_k=k^2+M^2 \; .
\end{equation}
We identify the dimensionless coupling constant $\lambda \equiv
  \frac{g^2}{(2 \, \pi   \, m)^2}$.

The contribution to the spectral density for $\omega^2 > W^2_k$
corresponds to the two particle cut of one massive and one
massless particle, and remains in the zero temperature limit. The
contribution  for $\omega^2< k^2$ corresponds to the Landau
damping cut, only present at finite temperature and below the
light cone.

This spectral density models all of the situations studied in the
previous sections:

\begin{itemize}
\item{ {\bf Case I:~$m \neq M$}.  In this case $\omega_k$  is away
  from the two particle threshold at
$W_k$ and above the light cone, thus away from the Landau damping
region. If $m<M$ the particle is stable and the self-energy
correction amounts to a renormalization of the dispersion
relation. If $m>M$ now $\omega_k$ lies in the two particle
continuum, and the particle pole moves off the physical sheet, in
this case the particle becomes a quasiparticle with a finite width
$\propto \lambda$. In both cases $\rho_k(\omega=\omega_k)\neq 0$
and $\rho_k(\omega) \neq 0$ for $\omega \gtrless \omega_k$. This
situation describes the case I analyzed in detail above and
corresponds to the most common cases of stable particles or
resonances.  }

\item{ {\bf Case II:~$m=M$}.  In this case the two particle
threshold
  is precisely at $\omega_k$, but the two particle
and Landau damping thresholds are well separated for any finite
momentum $k$. The behavior of the spectral density near the
threshold at $\omega_k$ is given by
\begin{equation}\label{caseIIspec}
 \rho_k(\omega)\buildrel{\omega \to \omega_k}\over =\lambda \; m^2 \,
\Bigg[\frac{2\omega_k}{m^2}(\omega-\omega_k)+\frac{T}{k}
\ln\left|\frac{\omega_k+k}{\omega_k-k} \right|\Bigg] \;
\Theta(\omega-\omega_k) .
\end{equation}
This spectral density is of the same form as that of case II given
by eq.(\ref{nT}). This case also applies to self energies of
charged particles in gauge theories where the emission or
absorption of massless gauge particles imply that $\omega_k$ is
actually the beginning of a branch cut in the spectral
representation of the charged particle propagator. While the
(instantaneous) Coulomb interaction is screened at finite
temperature by the polarization of the medium, transverse gluons
(in perturbation theory) and photons are only dynamically screened
by Landau damping\cite{pisa,blaizot,lebellac}. Thus the
self-energy of charged particles in general gauge theories
features the threshold infrared divergences associated with case
II and described by the threshold behavior given by eqs.(\ref{nT})
and (\ref{caseIIspec}). }

\item{{\bf Case III: $k >> m=M$, or $M=m=0$.} In the case $M=m$
and
  for a hard particle $k >>m$ or alternatively
for $M=m=0$, the behavior of the spectral density
eq.(\ref{specdensFinT}) above for $\omega_k \rightarrow k$ is seen
to be of the form
\begin{equation}
\rho_k(\omega)\buildrel{\omega \to k}\over = \lambda \; m^2 \;
\frac{T}{k} \; \ln\left|\frac{\mu(k)}{|\omega|-k}\right| \; ,
\end{equation}
\noindent where $\mu(k)= 2T \; \left(1 - e^{-\frac{k}{T}}\right)$.
In this case the Landau damping cut merges with the two particle
cut since $\omega_k=k$  and the spectral density features a finite
temperature logarithmic divergence at the position of the particle
mass shell. This is the situation described by case III analyzed
above and arises in the case of a hard charged particle, either
fermionic or bosonic in high temperature gauge theories, where
again the emission and absorption of transverse gauge fields,
which are only dynamically screened introduces logarithmic
divergences. In this case the notion of a damping rate which is
associated with exponential relaxation is not the correct one to
describe the relaxation of single quasiparticle excitations. }

\end{itemize}

\section{Time evolution in quantum mechanics}
While we have focused the discussion above on the evolution of the
expectation value of quantum fields, the DRG approach also offers
a systematic method to study the evolution of quantum states. In
particular, as explained below in detail, the DRG offers an
alternative to the adiabatic switching-on procedure to generate
\emph{exact} eigenstates of the interacting Hamiltonian from the
eigenstates of the non-interacting system. Along the way this DRG
formulation extracts the energy shifts, wave function
renormalization constants and offers a clear description of
Fermi's Golden rule.

Consider the time evolution of a quantum state in an interacting
theory with Hamiltonian
\begin{equation}\label{totalHam}
H=H_0+ \lambda \; H_I \; ,
\end{equation}
\noindent namely
\begin{equation}
|\psi(t)\rangle = e^{-iH(t-t_0)} ~|\psi(t_0)\rangle=e^{-iH_0 t} \;
U(t,t_0) \; e^{iH_0 t_0}|\psi(t_0)\rangle \; .
\end{equation}
Where we have introduced the unitary time evolution operator in
the interaction picture
\begin{equation}
U(t,t_0)= 1 -i \; \lambda \; \int^t_{t_0}H_I(t') \; dt'-\lambda^2
\; \int^t_{t_0}\int^{t'}_{t_0}H_I(t') \; H_I(t'') \; dt' \; dt'' +
{\cal O}(\lambda^3) ~~,~~ U(t_0,t_0)=1 \; , \label{unitop}
\end{equation}
\noindent with the interaction Hamiltonian in the interaction
picture
\begin{equation}
H_I(t)= e^{iH_0 t} \; H_I  \; e^{-iH_0 t} \; .
\end{equation}
It is convenient to pass to the interaction picture in which the
states are given by
\begin{equation}
|\psi(t)\rangle_i= e^{iH_0t} \; |\psi(t)\rangle \; ,
\end{equation}
\noindent and their time evolution is given by
\begin{equation}
|\psi(t)\rangle_i= U(t,t_0) \; |\psi(t_0)\rangle_i \; .
\end{equation}
Clearly, the time evolution of quantum states is an initial value
problem.

The adiabatic, or Gell-Mann-Low theorem asserts that introducing
the \emph{adiabatic} time evolution operator $U_{\epsilon}(t,t_0)$
by replacing
\begin{equation}
H_I(t) \rightarrow e^{-\epsilon |t|} \; H_I(t) \; ,
\end{equation}
\noindent in the time evolution operator $U(t,t_0)$ given by
(\ref{unitop}), the states
\begin{equation}
|\Psi_n\rangle = U_{\epsilon}(0,-\infty) \; |n\rangle \; ,
\end{equation}
\noindent constructed out of the  eigenstates $|n\rangle$ of the
\emph{non-interacting Hamiltonian}  $H_0$ with energy $E_n$, are
\emph{exact} eigenstates of the total interacting Hamiltonian $H$,
and the exact eigenvalues are then given by\cite{baymQM}
\begin{equation}
E_n+\Delta E_n =
\frac{\langle\Psi_n|H|\Psi_n\rangle}{\langle\Psi_n|\Psi_n\rangle}
\; .
\end{equation}
We now show how the dynamical renormalization group provides an
illuminating alternative to this procedure. To make this approach
more clear and to establish contact with the familiar results, we
now consider the case in which $|\psi(t_0)\rangle_i$ is an
eigenstate of the non-interacting Hamiltonian $H_0$, namely
\begin{equation}
|\psi(t_0)\rangle_i = |n\rangle ~~;~~ H_0 |n\rangle = E_n
|n\rangle \; .
\end{equation}
To highlight the main ideas of the DRG, we will focus on the
evolution of the (persistence) amplitude,
\begin{equation}
C_n(t)= \langle n|\psi(t)\rangle_i  \; ,
\end{equation}
\noindent however the time evolution of the off-diagonal overlaps
$C_m(t)= \langle m|\psi(t)\rangle_i; m\neq n$ can be studied along
the same lines.

Inserting the identity $\sum_m |m\rangle \langle m|=1$
appropriately, we find for the amplitude $C_n(t)$ the following
expression,
\begin{eqnarray}\label{Cnoft}
&& C_n(t) =  \langle n|U(t,t_0) \psi(t_0)\rangle_i =
C_n(t_0)\Bigg[1 - i\;
    \lambda \; (t-t_0) \;  \langle n|H_I|n\rangle
 -\frac{\lambda^2}{2} \; (t-t_0)^2 \;   \langle n|H_I|n\rangle^2
 \nonumber \\&& -\lambda^2 \; \int^t_{t_0} \int^{t'}_{t_0}\sum_{m\neq
n} |\langle n|H_I|m \rangle|^2 \;  e^{i(E_n-E_m)(t'-t'')} \; dt'
\; dt'' + {\cal O}(\lambda^3)\Bigg]  \; ,
\end{eqnarray}
\noindent where we have written $C_n(t_0)=\langle
n|\psi(t_0)\rangle_i$ despite our choice of initial state for
which $C_n(t_0)=1$, to emphasize that the initial amplitude
factors out.

It proves convenient to introduce the spectral density
\begin{equation}
\rho_n(\omega) =\sum_{m\neq n} |\langle n|H_I|m \rangle|^2
 \; \delta(\omega-E_m) \label{specQM}\; ,
\end{equation}
\noindent in terms of which and after the straightforward
integration over the time variables $t',t''$, eq.(\ref{Cnoft})
becomes
\begin{eqnarray}\label{Cnoft2}
&& C_n(t) =  C_n(t_0)\Bigg\{1 - i \; \lambda \; (t-t_0) \;
\langle
  n|H_I|n\rangle
 -\frac{\lambda^2}{2} \; (t-t_0)^2 \;  \langle n|H_I|n\rangle^2
 \\&& -\lambda^2 \; \int d\omega \;  \rho_n(\omega) \;
  \left[\frac{i}{E_n-\omega}
 \left(t-t_0-\frac{\sin[(E_n-\omega)(t-t_0)]}{E_n-\omega} \right)+
  \frac{1-\cos[(E_n-\omega)(t-t_0)]}{(E_n-\omega)^2}\right]+{\cal O}(\lambda^3)
 \Bigg\}\; .\nonumber
\end{eqnarray}
Using formulae (\ref{formula1}) and (\ref{formula5}) of appendix
\ref{appendix:formulae} we find the asymptotic long time limit to
be given by,
\begin{eqnarray}\label{Cnoftt}
 C_n(t) = && C_n(t_0)\Bigg\{1 - i\; \lambda \; (t-t_0) \;  \langle
 n|H_I|n\rangle
 -\frac{\lambda^2}{2} \; (t-t_0)^2 \;  \langle n|H_I|n\rangle^2
 -i(t-t_0) \; \lambda^2 \; {\sum_m}'~ \frac{|\langle n|H_I|m
 \rangle|^2}{E_n-E_m}\nonumber \\ && -\lambda^2 \; \pi \;  (t-t_0) \;
 \rho_n(E_n)-\lambda^2 \; {\sum_m}'~ \frac{|\langle n|H_I|m
 \rangle|^2}{(E_n-E_m)^2}+ \mathcal{O} \left( \frac{1}{t-t_0},\lambda^3
 \right)\Bigg\} \; . \end{eqnarray}
\noindent where ${\sum_m}'$ refers to the sum over all the states
with $E_m \neq E_n$. We write the expression above as,
\begin{equation}
C_n(t)=C_n(t_0)\left[1+\lambda \;
\mathcal{S}_n^{(1)}(t)+\lambda^2 \;
\mathcal{S}_n^{(2)}(t)-\lambda^2 \;
\mathcal{Z}_n^{(2)}+\mathcal{O}\left( \frac{1}{t-t_0},\lambda^3
\right)\right] \label{Cnoftsec}\; ,
\end{equation}
\noindent where $\lambda^{(i)}\mathcal{S}_i^{(i)}(t)$ are secular
terms (linear and quadratic in time respectively), the  term
$-\lambda^2 \mathcal{Z}_n^{(2)}$ in eq.(\ref{Cnoftsec}) is the
time independent term in eq.(\ref{Cnoftt}).

Applying here the dynamical renormalization group, we introduce
the amplitude renormalization in the form
\begin{equation}
C_n(t_0)= C_n(\tau) \; R_n(\tau)~~;~~R_n(\tau)= 1+ \lambda \;
r_n^{(1)}(\tau)+\lambda^2 \;  r_n^{(2)}(\tau)+{\cal O}(\lambda^3)
\; . \label{renoamp}
\end{equation}
Hence,
\begin{equation}\label{cnrenor}
C_n(t)=C_n(\tau)\left\{1+\lambda
\left[r_n^{(1)}(\tau)+\mathcal{S}_n^{(1)}(t)\right]+\lambda^2
\left[r_n^{(2)}(\tau)+ \mathcal{S}_n^{(2)}(t)+r_n^{(1)}(\tau) \;
\mathcal{S}_n^{(1)}(t)\right]-\lambda^2 \;
\mathcal{Z}_n^{(2)}+{\cal
  O}(\lambda^3) \right\}
\end{equation}
The counterterms $r_n^{(i)}(\tau)$ are required to cancel the
secular terms at a time scale $t=\tau$, leading to
\begin{eqnarray}\label{relas}
r_n^{(1)}(\tau)& = & -\mathcal{S}_n^{(1)}(\tau)\quad , \quad
r_n^{(2)}(\tau) = -\mathcal{S}_n^{(2)}(\tau)+[r_n^{(1)}(\tau)]^2\;
.
\end{eqnarray}
The independence of the amplitude $C_n(t)$ on the arbitrary time
scale $\tau$, namely $dC_n(t)/d\tau =0$ leads to the dynamical
renormalization group equation
\begin{equation}
\dot{C}_n(\tau)\left\{1+\lambda \; \left[ r_n^{(1)}(\tau)+
\mathcal{S}_n^{(1)}(t)\right]+{\cal O}(\lambda^3)\right\}+
C_n(\tau)\left\{\lambda \dot{r}_n^{(1)}(\tau)+\lambda^2 \;
\left[\dot{r}_n^{(2)}(\tau)+ \dot{r}_n^{(1)}(\tau) \;
\mathcal{S}_n^{(1)}(t) \right]+{\cal O}(\lambda^3) \right\}=0
\end{equation}
\noindent where the dot stands for derivative with respect to
$\tau$. Keeping terms up to second order in $\lambda$ yields,
$$
\frac{\dot{C}_n(\tau)}{C_n(\tau)} = - \frac{d}{d\tau}\left\{
\lambda \; r_n^{(1)}(\tau) + \lambda^2 \; \left[ {r}_n^{(2)}(\tau)
-\frac12 \;
  \left( r_n^{(1)}(\tau) \right)^2 \right] \right \}+{\cal O}(\lambda^3)
$$
Notice that the $t$-dependent pieces cancelled out identically.
Such cancellation is necessary for the consistency of the DRG to
second order in $\lambda$ . We find using the counterterms given
by eq.(\ref{relas}), \be\label{solC} C_n(\tau) = C_n(t_0) \;
e^{\lambda \; {S}_n^{(1)}(\tau) + \lambda^2
  \;\left\{ \mathcal{S}_n^{(2)}(\tau) - \frac12 \; \left[
    {S}_n^{(1)}(\tau) \right] \right \} } \; .
\ee Using now the explicit form of the secular terms which are
read off from eq.(\ref{Cnoftt}), we finally find the DRG solution
to second order to be given by
\begin{equation}\label{solQS}
C_n(\tau)= C_n(0) \;  e^{-i \Delta E_n \tau} \; e^{-\frac{\Gamma_n
    \tau}{2}} \; ,
\end{equation}
\noindent with the energy shift and width up to
$\mathcal{O}(\lambda^2)$ given by
\begin{eqnarray}
\Delta E_n & = & \lambda \; \langle n|H_I| n\rangle + \lambda^2 \;
{\sum_m}'~ \frac{|\langle n|H_I|m
 \rangle|^2}{E_n-E_m} \label{energyshift} \quad , \quad
 \Gamma_n =  2 \, \pi \; \rho_n(E_n)  \; .
 \end{eqnarray}
Notice that the terms in $\tau^2$ cancel identically  in the
exponent of eq.(\ref{solC}) and (\ref{solQS}). Again, such
cancellation is necessary for the consistency of the DRG.

Finally, inserting eq.(\ref{solQS}) into  eq.(\ref{cnrenor}) and
choosing the arbitrary renormalization scale $\tau$ to coincide
with the time $t$, we find the asymptotic long time behavior of
the dynamical renormalization group resummed amplitude as
\begin{equation}
C_n(t) \buildrel{t \to \infty}\over= \mathcal{Z}_n \;  e^{-i
\Delta
  E_n t} \;  e^{-\frac{\Gamma_n t}{2}} \; .
\end{equation}
The energy shift $\Delta E_n$ is clearly the same obtained in
familiar perturbation theory, the decay rate $\Gamma_n$ coincides
with the transition probability per unit time obtained from
Fermi's Golden rule, and the wave function renormalization
\begin{equation}
\mathcal{Z}_n = 1-\lambda^2 \;{\sum_m}'~ \frac{|\langle n|H_I|m
 \rangle|^2}{(E_n-E_m)^2} = \frac{\partial}{\partial
 E_n}\left(E_n+\Delta E_n\right)\label{ZQS} \; ,
 \end{equation}
\noindent is the same as obtained in usual perturbation theory and
determines the overlap between the unperturbed state and the
\emph{exact} eigenstate of the Hamiltonian\cite{baymQM}. Thus, we
see that the dynamical renormalization group leads to an
alternative formulation of the construction of the exact
eigenstates which allows to extract the energy shifts, widths and
weights and that coincides with the familiar setting of quantum
mechanics.

There is an important byproduct of this exercise: by considering
the perturbative expansion up to second order in the interaction
Hamiltonian, we found secular terms that contain the square of the
first order secular terms. These must cancel consistently and
systematically in the final DRG equation since they are accounted
for in the resummation furnished by the DRG. Indeed we found such
cancellation in eqs.(\ref{solC}) and (\ref{solQS}). This is the
equivalent of the renormalizability of the perturbative expansion.
Thus, this example not only provides a pedagogical framework to
explore and confirm the dynamical renormalization group, but also
manifestly shows the renormalizability in the sense that higher
order terms that appear in the perturbative expansion, which are
associated with the expansion of first order terms in the
solution, cancel systematically in the dynamical renormalization
group equation.

\section{Dynamical Renormalization Group: resummations}

The examples studied in detail above clearly indicate that the
dynamical renormalization group provides a \emph{resummation} of
the perturbative series. Thus the following question arises :
which type of resummation is implemented by the DRG?. The answer
to this important question can be gleaned from the examples
corresponding to the different cases discussed in the previous
sections.

\begin{itemize}

\item{{\bf Case I:} in the case in which the position of the
particle
  mass shell near
$\omega_k$ is far away from the multiparticle thresholds in the
perturbative expansion, then the DRG provides a resummation which
is akin to the geometric Dyson sum of the propagator with the
perturbative self-energy but keeping only the \emph{secular terms}
at each order. The DRG yields the asymptotic long time behavior of
the expectation value of the field in the case of stable particles
or unstable quasiparticles with a
 lifetime (width or decay rate) determined by the spectral density at
  the position of the mass shell of the particle.}

 \item{ {\bf Cases II and III:} In the case in which there are
 threshold divergences in the propagator corresponding to
 the case when the position of the particle mass shell at
 $\omega=\omega_k$ coincides with the beginning (or end point)
 of a branch cut, the DRG provides a resummation of the
 \emph{self-energy kernel} along with a resummation of the
 perturbative series for the propagator \textit{ \`a la} Dyson. That
 this is indeed the case can be understood from the study in section
 \ref{sec:BN} where
 Bloch-Nordsieck resummation of the propagator exponentiates
 the logarithmic threshold divergences. This resummation
 of the propagator is akin to that via the usual Euclidean
 renormalization group and leads to a power law
 relaxation, just as the scaling behavior of renormalization group
 improved correlation functions. The case of
 threshold divergences at \emph{finite temperature} is much more
 severe by the enhancement of the infrared behavior by
 the Bose-Einstein occupation factors. In this case there is no
 readily available Bloch-Nordsieck or renormalization
 group resummation for the Euclidean propagator, but the DRG furnishes
 the resummation directly in real time.
 Thus, in conclusion, in all of these cases in which there are
 infrared divergences, typically arising from the emission
 and absorption of massless quanta, the DRG provides a resummation of
 the self-energy kernel along with the
 Dyson type resummation of the perturbative series.

 It is important to highlight that the DRG does all of these
 resummations automatically. Whereas in the usual approach,
 there is first the resummation of the Dyson series of the propagator,
 and in the case of threshold divergences
  the self-energy has to be further resummed independently, either by the
 Bloch-Nordsieck method or alternatively by
  the Euclidean renormalization group. Thus, insofar as extracting the
 asymptotic long time behavior of the expectation
  values, the DRG provides a powerful, systematic method with a
 straightforward implementation which furnishes several
  stages of resummation within a single approach.  }

\end{itemize}

\vspace{2mm}

{\bf Secular terms and separation of time scales:} For any finite
time the perturbative expansion is well defined and is free of any
divergences (ultraviolet divergences had been properly
renormalized). It is for long times that secular terms that grow
in time  invalidate the perturbative expansion. Consider the
example of case I where the solution to first order in the
coupling is given by eq.(\ref{secular1}). It is clear from this
expression that the perturbation expansion breaks down at a time
scale $t_b$ when
\begin{equation}
\lambda~t_b \left|   \frac{i
\,\mathrm{Re}\Sigma^{(1)}_k(\omega_k)}{2\omega_k}-
\frac{\pi \; \rho^{(1)}_k(\omega_k)}{2\omega_k}\right|\sim 1
\; .
\end{equation}
This time scale is obviously of the same order as the relaxation
time $t_b \simeq t_{rel}=\Gamma^{-1}_k$ where $\Gamma_k$ is the
width given by eq.(\ref{masswidth}). Secular terms dominate the
integral in eq.(\ref{general}) for sufficiently late times. Let
$t_{\rho}$ be such time scale, namely, for $ t \geq t_{\rho}$ we
can approximate the integral in eq.(\ref{general}) by the
corresponding secular terms. As is clear from the explicit
examples discussed above, the nature of the secular terms as well
as the scale $t_{\rho}$ depend on the behavior of the spectral
density  near $ \omega = \omega_k $. The detailed behavior of the
spectral density depends  both on the field theory studied as well
as  the properties of the medium or the initial state chosen.

Thus, the perturbative expansion features secular terms but it is
still reliable in a time interval such that
\begin{equation}\label{separation}
t_{\rho} \ll t \ll t_{rel} \; .
\end{equation}
Therefore, if there is a wide separation of time scales such that
$t_{rel} \gg t_{\rho}$, then, there is a wide region in time
during which secular terms can be clearly identified and yet the
perturbative expansion  is valid. The secular terms can be
extracted and renormalized during this large interval and the DRG
resummation extends the result uniformly for $t \gg t_{rel}$,
namely beyond the validity of the original perturbative expansion.

Since the spectral densities do not depend on the coupling (to
lowest order) but the relaxation time scale is such that $t_{rel}
\sim 1/[m \, \lambda]$ a wide separation of time scales is
justified in the weak coupling limit. In higher orders the width
is always at least one more power of the coupling as compared to
the spectral density. Hence, if there is a separation of scales in
lowest order, such separation will remain in higher orders.

Thus, the conclusion of this discussion is that the DRG allows a
systematic improvement of the perturbative series, leading to a
uniform expansion,  in a weakly coupled theory in which  a wide
separation between the typical time scale beyond which there
emerge secular terms and the relaxation time scale is justified.

\bigskip

{\bf DRG and Euclidean RG:}

\bigskip

The connection with the more familiar Euclidean renormalization
group can be established with the following example. Consider a
theory of a single scalar field $\phi$ with a quartic
self-interaction $\mathcal{L}_{int} = -\frac{\lambda_0}{4!}
\phi^4$ in four space-time dimensions defined with an upper
momentum cutoff $\Lambda$. The $2\rightarrow 2 $ particle
scattering amplitude with all equal external momentum  is given by
\begin{equation}
\Gamma^{(4)}(p,p,p,p)=\lambda_0 - \frac{3}{2} \; \lambda^2_0 \;  t
\quad , \quad t = \ln\Big ( \frac{\Lambda}{p} \Big)\; .
\end{equation}
Thus we see that the usual logarithmic divergences in the
scattering amplitude can be interpreted as \emph{secular terms} in
the perturbative expansion.

Introducing $\lambda_R[\kappa]$, the coupling renormalized at a
scale $\kappa$  as
\begin{equation}
\lambda_0 = \lambda_R[\kappa] \;
Z_{\lambda}[\kappa]~~;~~Z_{\lambda}[\kappa]=1+\lambda_R
 \; z_1[\kappa] +\mathcal{O}(\lambda^2) \; ,
\end{equation}
\noindent and choosing $z_1[\kappa]=\ln(
\frac{\Lambda}{\kappa})=\tau$ to cancel the secular term in the
variable $t$ in the scattering amplitude at the scale $p=\kappa$,
we find to second order in the renormalized coupling
\begin{equation}
\Gamma^{(4)}(p,p,p,p)=\lambda_R[\kappa] + \frac{3}{2} \;
\lambda^2_R[\kappa] \;  (\tau-t) \quad , \quad t = \ln\Big (
\frac{\Lambda}{p} \Big)  \quad , \quad \tau = \ln\Big (
\frac{\Lambda}{\kappa} \Big) \; .
\end{equation}
Since the scattering amplitude cannot depend on the arbitrary
renormalization scale $\kappa$, namely
\begin{equation}
\frac{d}{d\tau}\Gamma^{(4)}(p,p,p,p)=0 \; ,
\end{equation}
\noindent the renormalization group equation for the coupling
follows, namely
\begin{equation}
\dot{\lambda}_R+ \frac{3}{2} \; \lambda^2_R =0 \Rightarrow
\beta_{\lambda} = -\frac{3}{2} \; \lambda^2_R ~~;~~
\beta_{\lambda}= \kappa  \; \frac{\partial}{\partial \kappa}
\lambda_R[\kappa] \; .
\end{equation}
Solving this renormalization group equation, where
$\beta_{\lambda}$ is recognized as the usual renormalization group
beta function, and choosing the scale $\tau$ to coincide with $t$,
we obtain the renormalization group improved scattering amplitude
\begin{equation}
\Gamma^{(4)}(p,p,p,p) = \frac{\lambda_R[\kappa]}{1+\frac{3}{2} \;
\lambda_R[\kappa] \;
  \ln\Big(\frac{\kappa}{p} \Big)} \; .
\end{equation}
This expression reveals that the renormalization group is
resumming the geometric series of one loop bubbles but only
keeping the secular terms in the loops, namely the logarithmic
terms. This resummation is  tantamount to summing the leading
logarithms.

\section{Summary and conclusions}

In this article we studied the real time evolution and relaxation
of expectation values of quantum fields as well as the evolution
of the states in  quantum mechanics by implementing the dynamical
renormalization group.

The time evolution of expectation values of a bosonic quantum
field is studied as an initial value problem. The fully
renormalized equation of motion for the expectation value is
obtained in linear response to an external current that is
adiabatically switched on. We first studied the time evolution in
terms of the Laplace transform of the equation of motion with a
one-loop retarded self-energy, which is tantamount to a Dyson
resummation (geometric series) of the self-energy. The real time
evolution was obtained  in several cases of interest: stable
particles and quasiparticles (resonances) and in the case in which
the propagator features  threshold infrared divergences, which are
a consequence of absorption and emission of massless quanta at
zero or non-zero temperature. These threshold divergences
invalidate the perturbative expansion and are ubiquitous in gauge
theories. A Bloch-Nordsieck resummation leads to anomalous
relaxation in the case of logarithmic infrared singularities but
is not readily available in the case of more severe infrared
singularities.

A strict perturbative expansion of the solution of the equations
of motion (\ref{eqnofmotionscalar}) features secular terms, namely
terms that grow in time and invalidate the perturbative expansion.
We straightforwardly obtain the perturbative solution of the
equations of motion by expanding the Laplace transform solution
eqs.(\ref{sols})-(\ref{invlap}) in powers of $\lambda$.  One can
alternatively solve by iteration the evolution equations
(\ref{eqnofmotionscalar})\cite{drg1}.

We then implemented the dynamical renormalization group (DRG) to
resum these secular terms in all cases. The DRG effectively {\bf
resums the self-energy kernel} of the evolution equations
(\ref{eqnofmotionscalar}) while the Laplace transform
eqs.(\ref{sols})-(\ref{invlap}) provides the exact solution with
the input kernel (which is to first order in $\lambda$ here.) The
DRG method provides a consistent and systematic approach that at
once resums the secular terms. The DRG solution
eq.(\ref{finDRGsol}) is a {\bf uniform} asymptotic expansion for
late times.

The secular terms are linear in time in the case of stable
particles or quasiparticles and logarithmic in time or linear
times logarithmic in the case of threshold infrared divergences.
In the case of stable particles and quasiparticles (resonances)
the resummation implied by the DRG is the same as that of the
Dyson series or Laplace transform. However,  in the case of
threshold infrared divergences the DRG provides at once a
Bloch-Nordsieck type of resummation of the self-energy kernel {\bf
and} a Dyson series resummation of the solution.

We then studied the time evolution in quantum mechanics as an
initial value problem for the states by implementing the dynamical
renormalization group. Again, the perturbative expansion features
secular terms, which upon resummation via the DRG allow to extract
the energy shift and width of quantum levels as well as the wave
function renormalizations. Thus, the DRG provides a framework to
systematically obtain energy shifts and width of quantum states.
The consistency of the DRG to the second order of perturbations is
proved here. Such consistency is analogous to the
renormalizability of quantum field theory for the renormalization
group.

While we discussed primarily the case of bosonic theories, the
method can be simply generalized to fermionic and gauge
theories\cite{drg1,drgqed}.

There are many distinct advantages to this method that make it a
very powerful way to study relaxation \emph{directly in real
time}. While the most often used description of relaxation is
based on  the concept of quasiparticle widths and is associated
with exponential relaxation, the DRG does not bias the
description: if the secular terms are linear in time, indeed the
concept of a quasiparticle width is the relevant one. On the other
hand if the secular terms are more complicated, for example
logarithmic or linear in time times logarithmic, the concept of a
quasiparticle width and exponential relaxation is not the proper
one. The dynamical renormalization group leads to the correct
relaxational dynamics either power laws in the case of logarithmic
secular terms or more complicated exponentials in the case of
linear times logarithmic. These relaxational dynamics simply
cannot be extracted from the usual concept of quasiparticle widths
which are infrared divergent\cite{pisa,blaizot}.

Thus, one of the main conclusions of this article is that the
dynamical renormalization group provides a powerful, systematic
and easy to implement framework to study the relaxational dynamics
of expectation values and the time evolution of quantum states
directly in real time. This method implements a resummation of the
perturbative expansion which yields a uniform asymptotic long time
behavior and reveals the correct relaxational dynamics, all within
one simple method.

While we have illustrated the main aspects of the DRG to lowest
order in the perturbative expansion, the next step is to extend
the DRG calculations to higher loops in field theory.

\section{Acknowledgments}

We thank Emil Mottola for probing questions and useful
discussions. The work of D.B.\ was supported in part by the US
National Science Foundation under grants PHY-9988720 and
NSF-INT-9905954.

\appendix

\section{\label{sec:appendix1} Renormalized equation of motion and its
Laplace transform}

The wave function and mass renormalization [eq.(\ref{contra})] are
defined by the counterterms  $\delta m^2$ and $Z$ which  are
consistently computed in a perturbative expansion in terms of the
coupling constant.  Hence, the renormalized retarded self-energy is related to
the bare one by
\begin{equation}\label{Pi}
\widetilde{\Sigma}_k(s)\equiv (Z-1)(s^2+\omega^2_k)
+\widetilde{\Sigma}_{k,bare}(s) + \delta m^2 \; ,
\end{equation}
\noindent Using this renormalized self-energy the tadpole condition
eq.(\ref{tadpole}) leads to the equation of motion (\ref{eqnofmotionscalar}).

We perform now the  Laplace transform of eq.(\ref{eqnofmotionscalar}). We find,
\begin{equation}
\int_0^{\infty} dt \; e^{-st} \; {\ddot\phi}_k(t) = s^2 \,
\varphi_k(s) - {\dot \phi}_k(0) -s \, \phi_k(0) \; .
\end{equation}
In the Laplace transform of the kernel term  we change the
integration variable $ t' $ by  $ x \equiv t-t'>0$ and then the
integration variable $ t $ by $ y \equiv t-x $. We obtain,
\begin{eqnarray}
&&\int_0^{\infty} dt \; e^{-st} \; \int_{-\infty}^t dt' \;
{\Sigma}_k(t-t') \;   { \phi}_k (t') = \int_0^{\infty} dt \;
e^{-st} \;  \int_0^{\infty} dx \; { \Sigma}_k(x) \; {\phi}_k(t-x)
= \cr \cr &&=\int_0^{\infty} dx \; e^{-sx} \; { \Sigma}_k(x)
\int_{-x}^{\infty} dy \; e^{-sy} \; { \phi}_k( y) =\int_0^{\infty}
dx \; e^{-sx} \; { \Sigma}_k(x) \left[ \frac{ e^{sx} -1}{s} \; {
\phi}_k(0) + \int_0^{\infty} dy \; e^{-sy} \; {\phi}_k( y)\right]
=\cr \cr && =\frac{ { \phi}_k( 0)}{s} \left[ {\tilde\Sigma}_{ k}
(0)-{\tilde \Sigma}_{k} (s) \right] + {\tilde \Sigma}_{ k} (s) \;
\varphi_k(s) \; ,
\end{eqnarray}
where we used that $  {\phi}_k(t) =  {\phi}_k(0) $ for $ t < 0 $.

\section{\bf Asymptotic behavior of spectral
integrals}\label{appendix:formulae}

We summarize in this appendix the late time behavior of integrals
over the density of states used in Sections III.

In the formulas below $ p(y) $  stands for a smooth function for $
0 \leq y \leq \infty $. $ p(0)$ and $ p'(0) $  are finite. For
large $ y \; , p(y) $  are assumed to fall so that the integrals
over $ y $ converge at infinity.
\begin{eqnarray}
\int_{-A}^{\infty} \frac{dy}{y^2} \left( 1 - \cos yt \right) \;
p(y) &\buildrel{t \to \infty}\over=& \pi \; t \; p(0) + {\cal P}
\int_{-A}^{\infty} \frac{dy}{y^2} \;  p(y)  +  {\cal
O}\left( \frac{1}{t} \right) \; \label{formula1}\\
 \int_0^{\infty} \frac{dy}{y}
\left( 1 - \cos yt \right) \; p(y) &\buildrel{t \to \infty}\over=&
p(0) \left[ \ln(\mu \, t) + \gamma \right] + \int_0^{\infty}
\frac{dy}{y}\left[  p(y) -  p(0) \; \theta(\mu - y ) \right] +
{\cal O}\left( \frac{1}{t} \right)  \label{formula2}\\
\int_{-A}^{\infty} \frac{dy}{y^2} \left( 1 - \cos yt \right) \;
 { p}(y)\; \ln\frac{|y|}{\mu}
&\buildrel{t \to \infty}\over=& \pi \; t \; p(0)  \left[ 1 -
\gamma - \ln(\mu\, t) \right]  + \int_{-A}^{\infty} \frac{dy}{y^2}
\;  p(y)  \; \ln\frac{|y|}{\mu} +  {\cal O}\left( \frac{1}{t}
\right)
\label{formula3}\\
\int_{-A}^{\infty} \frac{ dy}{ y} \left( t - \frac{\sin yt}{y}
\right) \; p(y) \; \ln\frac{|y|}{\mu} &\buildrel{t \to
\infty}\over=& t \; {\mathcal P} \int_{-A}^{\infty} \frac{ dy}{ y}
\; p(y)\ln\frac{|y|}{\mu} \,-\pi \, { p}'(0)\left[\ln(\mu \, t  )
+ \gamma\right] + {\cal O}\left(
\frac{1}{t}   \right)\label{formula4}\\
\int_{-A}^{\infty} \frac{ dy}{y} \left( t - \frac{\sin yt}{y
}\right) \; p(y) &\buildrel{t \to  \infty}\over=& t \; {\mathcal
P} \int_{-A}^{\infty} \frac{ dy}{ y} \; p(y) - \pi \, p'(0) +
{\cal
O}\left( \frac{1}{t }  \right)\label{formula5}\\
\int_0^{\infty} \frac{dy}{y^2} \left( 1 - \cos yt \right) \; p(y)
&\buildrel{t \to \infty}\over=& \frac{\pi}2 \; t \; p(0) + p'(0)
\; \left[ \ln(\mu \, t) + \gamma \right] \cr \cr &+&
\int_0^{\infty} \frac{dy}{y^2} \left[  p(y) -  p(0) - y  \; p'(0)
\;\theta(\mu - y ) \right] + {\cal O}\left( \frac{1}{t} \right)
\label{formula6}\\
\int_0^{\infty} \frac{dy}{y} \left( t - \frac{\sin yt}{y}\right)
\; p(y) &\buildrel{t \to \infty}\over=& t \; p(0) \left[ \ln(\mu
\, t) + \gamma - 1 \right] + t \int_0^{\infty} \frac{dy}{y} \left[
p(y) -  p(0) \; \theta(\mu - y ) \right] + {\cal O}\left(
\frac{1}{t}   \right) \label{formula7}
\end{eqnarray}
where $ A $ is a fixed positive number and $\gamma = 0.5772157
\ldots$ is  Euler's constant and $\mathcal{P}$ stands for the
principal part.

Notice that the formulas are {\bf independent} of the scale $ \mu
$ as one can easily see since  the derivative with respect to $
\mu $ of the r. h. s. identically vanishes. The scale $ \mu $ has
been introduced just to have a dimensionless argument in the logs.

\section{\bf The Dynamical Renormalization Group in a Solvable Example}

In this appendix we present the dynamical renormalization group
through a simple example chosen for pedagogical reasons  which
allows to illustrate the fundamental features within a simple
setting.

We consider a damped harmonic oscillator:
\begin{equation}\label{oscamor}
\ddot{y}+y=-\epsilon \; \dot{y}~ \; , \epsilon \ll 1\; .
\end{equation}
This equation admits as exact solution,
\begin{equation}\label{solex}
y(t) = a \; e^{-\frac{\epsilon}{2} \, t } \; \cos\left( t \,
\sqrt{1 - \frac{\epsilon^2}{4} } + b \right)
\end{equation}
where the constants $a$ and $b$ are determined by the initial
conditions.

We seek a solution as a perturbative expansion in $\epsilon$, of
the form
$$
y(t)=y_0(t)+\epsilon \; y_1(t) + \epsilon^2 \; y_2(t) + {\cal
O}(\epsilon^3)
$$
where the $y_i(t)$ are solutions to the following hierarchy of
equations:
\begin{eqnarray}
&&\ddot{y}_0(t)+y_0(t) = 0 \quad , \quad \ddot{y}_1(t)+y_1(t) =
-\dot{y}_0(t)\quad , \quad \ddot{y}_2(t)+y_2(t) =
-\dot{y}_1(t)\quad , \quad  \cdots
\end{eqnarray}
These equations can be solved iteratively by starting from the
zero order solution
$$
y_0(t)= A\; e^{it} + \mbox{c.c} \; ,
$$
using the retarded Green's function
$$
G_{\text{ret}}(t)= \theta(t) \;  \sin t \; ,
$$
which is the solution of differential equation
$$
\left[ \frac{d^2}{dt^2} + 1 \right]G_{\text{ret}}(t)= \delta(t) \;
.
$$
Up to second order in $\epsilon$, the perturbative expansion of
the solution is given by
\begin{eqnarray}\label{solper}
y(t)&=&A\; e^{it} \left[1-\frac{\epsilon}{2} \;
t+\frac{\epsilon^2}{8} \; t^2+i \; \frac{\epsilon^2}{8} \;
t\right]+ \mbox{c.c.} +\mbox{non-secular} \nonumber
\end{eqnarray}

Note that this solution contains secular terms that grow in $t$,
the terms denoted by {\em non-secular} remain finite at all times.
We see that this asymptotic  expansion is only valid  for $
\epsilon \, t \ll 1 $.

The dynamical renormalization is achieved by introducing the
renormalization constant $Z(\tau)$ in the form
$$A=A(\tau)\;Z(\tau) $$
 with the expansion
$$
Z(\tau)=1+\epsilon \; z_1(\tau) +\epsilon^2 \; z_2(\tau) + {\cal
O}(\epsilon^3)
$$
The coefficients  $z_i(\tau)$ are chosen  to cancel the secular
terms at a the time scale $t=\tau$.  This is similar to choosing
the wave function renormalization in field theory to absorb the UV
(or IR)  divergences at a given momentum scale. Up to ${\cal
O}(\epsilon^2)$ we find
$$
z_1(\tau) = \frac{\tau}{2} \quad ; \quad z_2(\tau) =
\frac{\tau^2}{8}-i \; \frac{\tau}{8} \;.
$$
After this renormalization the solution (\ref{solper}) takes the
form,
\begin{eqnarray}\label{dampsolu}
y(t,\tau)&=&A(\tau)\; e^{it}\left[1-\frac{\epsilon}{2}\; (t-\tau)
+ \frac{\epsilon^2}{8}\;(t-\tau)^2 +i \;
\frac{\epsilon^2}{8}\;(t-\tau)\right]+
\mbox{c.c.}+\mbox{nonsecular}\;.
\end{eqnarray}
We have now an expansion in powers of $ \epsilon |t-\tau| $. This
asymptotic expansion is expected to be valid for $\epsilon
|t-\tau| \ll 1 $. That is, renormalization has  improved the
validity region of the solution from the neighborhood of $ t = 0 $
[see eq.(\ref{solper})] to the neighborhood of $ t = \tau $ which
is an arbitrary point.

Since $\tau$ is an arbitrary scale, the solution must not depend
on it. We therefore request
\begin{equation}\label{drg1}
\frac{dy(t,\tau)}{d\tau}  =0 \;.
\end{equation}
This condition leads to the following equation to this order
\begin{equation}\label{drg2}
\frac{\partial A(\tau)}{\partial \tau} +\left(\frac{\epsilon}{2}-i
\; \frac{\epsilon^2}{8}\right)A(\tau)=0 \;.
\end{equation}
where we have neglected terms of order higher than $ \epsilon^2$.
Eq.(\ref{drg2}) is the dynamical renormalization group equation to
this order. Eq.(\ref{drg2}) can be easily solved with the solution
$$
A(\tau)= A(0) \;  e^{-\frac{\epsilon}{2}\tau + i
\frac{\epsilon^2}{8}\tau} \;.
$$
Since the solution $y(t,\tau)$ is independent of $ \tau $ thanks
to eq.(\ref{drg1}) we  finally set $t=\tau$ which yields,
\begin{equation}\label{drg3}
y(t) = A(0) \;  e^{-\frac{\epsilon}{2} t +i
(1-\frac{\epsilon^2}{8})t }+ \mbox{c.c.}
\end{equation}
Comparing eq.(\ref{drg3}) with eq.(\ref{solex}) shows that
eq.(\ref{drg3}) provides an {\bf uniform} approximation to the
exact solution (\ref{solex}) up to $ {\cal O}(\epsilon^3) $. That
is, the dynamical renormalization group solution (\ref{drg3})
differs from the exact solution (\ref{solex}) by  $ {\cal
O}(\epsilon^3) $ whatever is $t$.

The solution (\ref{drg3}) is nothing but an envelope of the family
of solutions  (\ref{dampsolu}) parametrized by $\tau$\cite{kuni}.
In other words, the $\tau$-family (\ref{dampsolu}) provides a
solution for $t$ near $\tau$ while eq.(\ref{drg3}) is a good
approximation valid for {\bf all} $t$.

Further  examples can be found in
refs.\cite{goldenfeld,kuni,otrosdrg}.

\end{document}